\documentclass[reprint,
superscriptaddress,amsmath,amssymb,aps, longbibliography
]{revtex4-2}

\usepackage{graphicx}
\usepackage{dcolumn}
\usepackage{bm}
\usepackage{siunitx}
\usepackage[noend]{algorithmic}
\usepackage{algorithm}
\usepackage[skins]{tcolorbox}
\usepackage{setspace}
\usepackage{blkarray}

\usepackage{xcolor}
\usepackage{soul}
\usepackage{systeme}
\usepackage{footmisc}

\usepackage{scalerel} 

\makeatletter
\def\maketag@@@#1{\hbox{\m@th\normalfont\normalsize#1}}
\makeatother

\usepackage{booktabs}
\usepackage{hyperref}
\usepackage[capitalize]{cleveref}
\usepackage[utf8]{inputenc}

\usepackage[lining,semibold]{libertine} 
\usepackage{amsthm}
\usepackage[libertine, cmintegrals, bigdelims, vvarbb]{newtxmath}

\usepackage{amsmath}
\usepackage{amsfonts}
\usepackage{mathrsfs}
\usepackage{gensymb}
\usepackage{bbm}
\usepackage{dsfont}

\usepackage{pgfplots}
\usepgfplotslibrary{colormaps}

\usepackage{kbordermatrix}
\renewcommand{\kbldelim}{(}
\renewcommand{\kbrdelim}{)}

\usepackage{chemformula}
\usepackage[caption=false]{subfig}

\usepackage{soul}
\usepackage{xcolor}

\theoremstyle{definition}

\usepackage{scalerel} 

\definecolor{webgreen}{rgb}{0,.5,0}
\definecolor{webbrown}{rgb}{.6,0,0}
\definecolor{grigio}{rgb}{.85,.85,.85} 
\definecolor{RoyalBlue}{rgb}{0.0, 0.14, 0.4}
\definecolor{skyblue1}{rgb}{0.45,0.62,0.81}
\definecolor{skyblue2}{rgb}{0.2,0.39,0.64}
\definecolor{skyblue3}{rgb}{0.13,0.29,0.53}
\definecolor{scarlet1}{rgb}{0.93,0.16,0.16}
\definecolor{scarlet2}{rgb}{0.8,0,0}
\definecolor{scarlet3}{rgb}{0.64,0,0}

\definecolor{g}{gray}{0.50}

\hypersetup{%
    colorlinks=true, linktocpage=true, pdfstartpage=1, pdfstartview=FitV,%
    breaklinks=true, pdfpagemode=UseNone, pageanchor=true, pdfpagemode=UseOutlines,%
    plainpages=false, bookmarksnumbered, bookmarksopen=true, bookmarksopenlevel=1,%
    hypertexnames=true, pdfhighlight=/O,
    urlcolor=webbrown, linkcolor=RoyalBlue, citecolor=webgreen, 
    pdftitle={},%
    pdfauthor={Timur Aslyamov},%
    pdfsubject={},%
    pdfkeywords={},%
    pdfcreator={pdfLaTeX},%
    pdfproducer={LaTeX REVTeX}%
}

\begin{document}
\title{Nonequilibrium fluctuation-response relations for state observables}

\author{Krzysztof Ptaszy\'{n}ski}
\email{krzysztof.ptaszynski@ifmpan.poznan.pl}
\affiliation{Institute of Molecular Physics, Polish Academy of Sciences, Mariana Smoluchowskiego 17, 60-179 Pozna\'{n}, Poland}

\author{Timur Aslyamov}
\email{timur.aslyamov@uni.lu}
\affiliation{Complex Systems and Statistical Mechanics, Department of Physics and Materials Science, University of Luxembourg, 30 Avenue des Hauts-Fourneaux, L-4362 Esch-sur-Alzette, Luxembourg}

\author{Massimiliano Esposito}
\email{massimiliano.esposito@uni.lu}
\affiliation{Complex Systems and Statistical Mechanics, Department of Physics and Materials Science, University of Luxembourg, 30 Avenue des Hauts-Fourneaux, L-4362 Esch-sur-Alzette, Luxembourg}

\date{\today}

\begin{abstract}
Time-integrated state observables, which quantify the fraction of time spent by the system in a specific pool of states, are important in many fields, such as chemical sensing or the theory of fluorescence spectroscopy. We derive exact identities, called Fluctuation-Response Relations (FRRs), that connect the fluctuations of such observables to their response to external perturbations in nonequilibrium steady state of Markov jump processes. Using these results, we derive a first known upper bound on fluctuations of state observables, as well as some new lower bounds. We further demonstrate how our identities provide a deeper understanding of the mechanistic origin of fluctuations and reveal their properties dependent only on system topology, which may be relevant for model inference using measured data.
\end{abstract}

\maketitle

\section{Introduction}
The system responses to external perturbations and stochastic fluctuations of system observables are among the central topics of statistical physics. Close to equilibrium, the responses and fluctuations are strictly related via a seminal fluctuation-dissipation theorem~\cite{kubo1966fluctuation,kubo2012statistical,stratonovich2012nonlinear,marconi2008fluctuation,forastiere2022linear}. Far from equilibrium, such universal relations are no longer valid. However, in recent years, significant progress has been made in describing the universal properties of static responses to external perturbations in Markov processes or chemical reaction networks~\cite{lucarini2016response,santos2020response,falasco2019negative, mallory2020kinetic,owen2020universal,owen2023size,gabriela2023topologically,aslyamov2024nonequilibrium,aslyamov2024general,harunari2024mutual,cengio2025mutual,khodabandehlou2024affine,floyd2024learning,frezzato2024steady,floyd2024limits,gao2022thermodynamic,bao2024nonequilibrium,zheng2025spatial}. Similarly, research in recent decades has produced a wealth of universal laws describing the properties of fluctuations, such as fluctuation theorems~\cite{jarzynski1997nonequilibrium,crooks1999entropy,esposito2009nonequilibrium,seifert2012stochastic}, generalizations of the fluctuation-dissipation theorem~\cite{agarwal1972fluctuation,baiesi2009fluctuations,seifert2010fluctuation,prost2009generalized,altaner2016fluctuation,maes2020response,chun2023trade,shiraishi2023introduction,gao2024thermodynamic,tesser2024out}, or Thermodynamic and Kinetic Uncertainty Relations (TURs and KURs), lower bounding the current fluctuations in terms of average current and entropy production or traffic (also called activity), a quantity that measures the total number of
transitions per unit time in the system~\cite{barato2015thermodynamic,gingrich2016dissipation,pietzonka2016universal, pietzonka2017finite,horowitz2017proof,falasco2020unifying,horowitz2020thermodynamic,vu2020entropy,van2023thermodynamic,di2018kinetic,shiraishi2021optimal}.

Very recently, significant developments have also been made in connecting responses and fluctuations, in the spirit of the original fluctuation-dissipation theorem, in systems arbitrarily far from equilibrium~\cite{dechant2019arxiv,dechant2020fluctuation,falasco2022beyond,monnai2024kinetic}. In particular, Ref.~\cite{zheng2024information} used information theory to obtain an inequality bounding the precision of response (i.e., its ratio to fluctuations) of an arbitrary trajectory observable in Markov jump processes by the traffic. This result was later generalized to discrete-time Markov processes~\cite{liu2024dynamical} and open quantum systems~\cite{kwon2024fluctuation,van2024fundamental,liu2025response}. Complementing these results, Ref.~\cite{ptaszynski2024dissipation} obtained a similar Response-TUR (R-TUR), bounding the precision of response to kinetic perturbations (i.e., perturbations affecting the system kinetics, but not thermodynamic forces) in a Nonequilibrium Steady State (NESS) by the entropy production. This result was later proven in Ref.~\cite{aslyamov2024frr}, which further derived exact identities, called Fluctuation-Response Relations (FRRs), relating the current fluctuations in NESS to static responses to kinetic or thermodynamic perturbations. Subsequently, R-TUR has been generalized to the transient regime using information theory~\cite{kwon2024fluctuation}.

The kinetic and thermodynamic bounds derived in Refs.~\cite{zheng2024information,kwon2024fluctuation,liu2024dynamical} were also applicable to the precision of response of time-integrated state observables, which describe the fraction of time spent by the system in a given state. Such observables and their fluctuations have attracted significant attention in many fields, including diffusion~\cite{levy1940certains,dhar1999residence,godreche2001statistics,majumdar2002local,majumdar2007brownian,lapolla2018unfolding,lapolla2019manifestations,lapolla2020spectral}, motion of active particles~\cite{singh2019generalised,bresloff2020occupation}, theory of fluorescence microscopy~\cite{palo2006calculation,gopich2012theory}, optical systems~\cite{margolin2005nonergodicity,ramesh2024arcine}, nanoelectronics~\cite{utsumi2007full,utsumi2007fullb,utsumi2008full}, or chemical sensing~\cite{bialek2005physical,endres2008accuracy,mora2010limits,govern2012fundamental,govern2014energy,kaizu2014berg,lang2014thermodynamics,barato2015dispersion,ray2017dispersion,harvey2023universal}. In particular, in the latter context, they enter a famous Berg-Purcell bound for the precision of concentration measurements~\cite{berg1977physics}. Other types of inequalities bounding the fluctuations of state observables have also been derived in Ref.~\cite{koyuk2020thermodynamic} for the time-dependent scenario and in Ref.~\cite{macieszczak2024occupation} for the stationary case. However, the exact identities relating fluctuations and responses of state observables, analogous to the FRRs obtained for current~\cite{aslyamov2024frr}, have not been known so far. 

In our article, we close this gap by deriving exactly such identities [\cref{eq:covar-exact}]. Remarkably, we found that they are identical in structure to the FRRs proven for currents. In the companion paper~\cite{ptaszynski2025mixed}, we further show that FRRs with the same structure also apply to covariances of state and current observables. This is nontrivial since fluctuations and responses of state~\cite{bao2024nonequilibrium,owen2020universal,owen2023size,gabriela2023topologically,koyuk2020thermodynamic,macieszczak2024occupation} and current~\cite{koyuk2020thermodynamic,bakewell2023general,bao2024nonequilibrium,altaner2016fluctuation} observables often obey distinct physical laws (e.g., the standard TUR~\cite{barato2015thermodynamic,gingrich2016dissipation,pietzonka2016universal, pietzonka2017finite,horowitz2017proof,falasco2020unifying,horowitz2020thermodynamic,vu2020entropy,van2023thermodynamic,di2018kinetic,shiraishi2021optimal} does not apply to the former). Using our identities, we derive a first known upper bound on fluctuations of state observables [\cref{eq:bound-upper-var}]. We also obtain upper bounds for the precision of the equilibrium response to energy perturbations [Eqs.~\eqref{eq:vertex-ineq-dyn}--\eqref{eq:tur-vert}], and reveal a deep connection between previously unrelated results from Refs.~\cite{kwon2024fluctuation,zheng2024information, liu2024dynamical} and~\cite{macieszczak2024occupation}. Finally, we use a quantum dot example to show that FRRs provide insight into the origin of fluctuations by relating their behavior to the response properties of the system. 
In particular, in certain cases, one may associate the sign of covariances with the system topology, which may be relevant for model inference using measured data. 

\section{Framework}

\subsection{Setup}
We consider a continuous-time Markov jump process among $N$ discrete states. It is described by the graph whose nodes correspond to the system states and the undirected edges $e \in \mathcal{E}$ to the transitions between states. We further make the graph directed by assigning each edge $e$ a forward ($+e$) and reverse ($-e$) direction, so that the source of the directed edge $\pm e$, labeled $s(\pm e)$, is a target of the directed edge $\mp e$, labeled $t(\mp e)$. The transition rate associated with the directed edge $\pm e$ is denoted as $W_{\pm e}$. The steady state of the system is defined by
\begin{align}
\label{eq:NESS}
    d_t\boldsymbol{\pi} = \mathbb{W}\cdot\boldsymbol{\pi} = 0 \,,
\end{align}
where $\boldsymbol{\pi}=(\dots,\pi_n,\dots)^\intercal$ is the vector of state probabilities $\pi_n$ with $\sum_n\pi_n=1$. The matrix $\mathbb{W}$ is the rate matrix with off-diagonal elements $W_{nm}=\sum_{e}[ W_{+e}\delta_{s(+e)m}\delta_{t(+e)n} + W_{-e}\delta_{s(-e)m}\delta_{t(-e)n}]$, where $\sum_e$ denotes the summation over the undirected edges $e$, and the diagonal elements $W_{nn}=-\sum_{m \neq n}W_{mn}$.  We further employ a generic parameterization of the transition rates ~\cite{owen2020universal,aslyamov2024nonequilibrium}
\begin{align}
\label{eq:rates-model}
    W_{\pm e}=\exp(X_{\pm e}) \quad \text{with} \quad X_{\pm e}=V_{s(\pm e)}+B_e \pm S_e/2 \,,
\end{align}
where $B_e$ and $S_e$ are called symmetric and antisymmetric edge parameters, respectively, and $V_n$ are called vertex parameters. We note that this parametrization is not unique for a given Markov network. Consequently, these parameters cannot be unequivocally identified with the physical parameters of the system (e.g., energies or thermodynamic forces), but rather there is some freedom in relating them to physical parameters. For example, taking $V_n=0$, $B_e$ characterize the kinetic barriers, while $S_e$ is the change in entropy in the reservoir due to a transition $e$ that includes changes in thermodynamic forces and the energy landscape~\cite{rao2018conservation, falasco2023macroscopic, owen2020universal}. However, the energy landscape can be alternatively parametrized by vertex parameters $V_n$, which is particularly useful at equilibrium, as we will see below.

\subsection{State observables}
Our object of interest are the state observables that are time integrated along the stochastic trajectory of the system,
\begin{align}
\hat{o}(t) \equiv \frac{1}{t} \sum_{n} o_n \int_{0}^t \phi_n(t') dt' \,,
\end{align}
where $\boldsymbol{o} \equiv (\dots, o_n, \dots)^\intercal$ is the vector defining the observable, while $\phi_n(t)$ is the random variable taking the value $1$ when the state $n$ is occupied and $0$ otherwise. The average value of this observable is defined as
\begin{align}
\mathcal{O}(t) \equiv \langle \hat{o}(t) \rangle \,,
\end{align}
where $\langle \cdot \rangle$ denotes the average over the ensemble of stochastic trajectories. The covariance of two observables is in turn defined as
\begin{align}
\langle\!\langle \mathcal{O}(t),\mathcal{O}'(t) \rangle\!\rangle \equiv t \langle \Delta \hat{o}(t) \Delta \hat{o}'(t) \rangle \,,
\end{align}
where $\Delta \hat{o}(t) \equiv \hat{o}(t) - \langle \hat{o}(t) \rangle$. 

We further mostly focus on the long-time version of the quantities considered,
\begin{subequations}
    \begin{align}
\mathcal{O} &\equiv \lim_{t \rightarrow \infty} \mathcal{O}(t) = \sum_{n} o_n \pi_n \,, \\ \langle\!\langle \mathcal{O},\mathcal{O}' \rangle\!\rangle  &\equiv \lim_{t \rightarrow \infty} \langle\!\langle \mathcal{O}(t),\mathcal{O}'(t) \rangle\!\rangle \,.
\end{align}
\end{subequations}
We emphasize that while the average observable $\mathcal{O}$ is determined only by the stationary state $\boldsymbol{\pi}$, the covariances $\langle\!\langle \mathcal{O},\mathcal{O}' \rangle\!\rangle$ are dynamical quantities, dependent on the ensemble of stochastic trajectories. They are given by the algebraic expression
\begin{align}
\langle\!\langle \mathcal{O},\mathcal{O}' \rangle\!\rangle=\boldsymbol{o}^\intercal \mathbb{C} \boldsymbol{o}' \,,
\end{align}
where $\boldsymbol{o}' \equiv (\dots, o_n', \dots)^\intercal$, and $\mathbb{C}=[C_{mn}]$ is the covariance matrix of occupation times of different states. It is defined as
\begin{align}
C_{mn} \equiv \lim_{t \rightarrow \infty} \frac{1}{t} \langle \theta_m(t) \theta_n(t) \rangle \,,
\end{align}
where $\theta_n(t) \equiv \int_0^t [\phi_n(t')-\pi_n]dt'$. It can be calculated using the formula (see Appendix~\ref{app:equiv-covmat})~\cite{lapolla2018unfolding,lapolla2019manifestations,lapolla2020spectral,macieszczak2024occupation}
\begin{align} \label{eq:covmat-expr}
    \mathbb{C}=-\mathbb{W}^D \cdot \text{diag}(\boldsymbol{\pi})-[\mathbb{W}^D \cdot \text{diag}(\boldsymbol{\pi})]^\intercal \,,
\end{align}
where $\mathbb{W}^D$ is the Drazin inverse of the rate matrix, which has been generically defined in Ref.~\cite{drazin1958pseudo}. In our context, it corresponds to a unique solution of the equation $\mathbb{W} \mathbb{W}^D=\mathbb{W}^D \mathbb{W}=\mathds{1}-\boldsymbol{\pi} \boldsymbol{1}^\intercal$ where $\boldsymbol{1}=(\ldots,1,\ldots)^\intercal$. See Refs.~\cite{crook2018drazin,landi2023current,bao2024nonequilibrium,harvey2023universal} for more of its properties and applications for characterizing fluctuations and responses.

\subsection{Static responses}
We also consider the static responses of the state observables, that is, the linear response of the steady-state value of the observable $\mathcal{O}$ to some parameter $p$ that controls the transition rates $W_{\pm e}$~\cite{aslyamov2024general}. Operationally, this involves measuring the responses after a time interval following the perturbation of the parameter $p$ that is long enough for the system to relax to its new stationary state.
Throughout our paper, we focus on a situation where the vector $\boldsymbol{o}$ defining the observable does not depend on the parameter $p$. For such a case, the static response of the observable $\mathcal{O}$ reads
\begin{align}
d_{p} \mathcal{O}=\boldsymbol{o}^\intercal d_{p} \boldsymbol{\pi} \,,
\end{align}
where $d_{p} \boldsymbol{\pi}$ is the static response of the stationary probability vector. It can be calculated as~\cite{ptaszynski2024critical}
\begin{align} \label{eq:response-drazin}
d_{p} \boldsymbol{\pi}=-\mathbb{W}^D (d_{p} \mathbb{W}) \boldsymbol{\pi} \,.
\end{align}
This expression is rederived in Appendix~\ref{app:der-response}. We notice that the Drazin inverse form of \cref{eq:response-drazin} is an alternative to the method from Refs.~\cite{aslyamov2024nonequilibrium,aslyamov2024general}. We further note that when the vector $\boldsymbol{o}$ depends on the parameter $p$, the results presented later can be applied upon replacement $d_p \mathcal{O} \rightarrow d_p \mathcal{O}-\boldsymbol{\pi}^\intercal d_p \boldsymbol{o}$. We also emphasize that derivatives over the edge and vertex parameters, $d_p \mathcal{O}$ ($p \in \{ V_{n},B_e,S_e \}$), are independent of the particular parameterization of transition rates in terms of these parameters~\cite{owen2020universal}.

\section{Results}
\subsection{Fluctuation-response relations}
Let us now present our main result, namely, the exact identities relating the covariance matrix elements to static responses of state probabilities. They are proven in the Appendix~\ref{app:proof-frr}. They read as
\begin{subequations}
\label{eq:covmat-exact}
\begin{align} \label{eq:covmat-exact-gen}
C_{mn} &= \sum_{\pm e}\frac{1}{\varphi_{\pm e}}d_{X_{\pm e}} \pi_m d_{X_{\pm e}} \pi_n  \\
\label{eq:covmat-exact-sym}
 &= \sum_{e}\frac{\tau_e}{j_e^2}d_{B_e} \pi_m d_{B_e} \pi_n \\ \label{eq:covmat-exact-antisym}
 &= \sum_{e} \frac{4}{\tau_e} d_{S_e} \pi_m d_{S_e} \pi_n\,,
\end{align}
\end{subequations}
where $\varphi_{\pm e} \equiv W_{\pm e} \pi_{s (\pm e)}$ is the unidirectional probability flux along the edge $\pm e$, $j_e \equiv \varphi_{+e}-\varphi_{-e}$ is the oriented current along the edge $e$, and $\tau_e \equiv \varphi_{+e}+\varphi_{-e}$ is the unoriented traffic through that edge. Here, recall, $\sum_e$ denotes the summation over undirected edges $e$, while $\sum_{\pm e}$ denotes the summation over forward and backward directed edges. We note that Eq.~\eqref{eq:covmat-exact-sym} still holds at a stalling edge where both the denominator, $j_e^2$, and the numerator, $\tau_e d_{B_e} \pi_m d_{B_e} \pi_n$, tend to zero, because their ratio remains finite~\cite{aslyamov2024general,aslyamov2024frr}.  

The above identities produce analogous fluctuation-response relation (FRRs) for generic state observables,
\begin{subequations} \label{eq:covar-exact}
\begin{align} \label{eq:covar-exact-gen}\langle\!\langle\mathcal{O},\mathcal{O}'\rangle\!\rangle &=\sum_{\pm e} \frac{1}{\varphi_{\pm e}} d_{X_{\pm e}}\mathcal{O}d_{X_{\pm e}}\mathcal{O}' \\
\label{eq:covar-exact-sym}
 &= \sum_{e}\frac{\tau_e}{j_e^2}d_{B_e}\mathcal{O}d_{B_e}\mathcal{O}' \\
\label{eq:covar-exact-antisym}
 &= \sum_{e} \frac{4}{\tau_e} d_{S_e}\mathcal{O}d_{S_e}\mathcal{O}'\,.
\end{align}
\end{subequations}
The identities~\eqref{eq:covmat-exact} and~\eqref{eq:covar-exact}, which reveal an intimate link between fluctuations and responses, are the main result of our paper. We recall that the structure of these identities, expressing covariances of state observables in terms of products of their responses to perturbations of transition rates, is identical to the FRRs previously derived for currents~\cite{aslyamov2024frr}, and generalized to covariances of state and current observables in a companion paper~\cite{ptaszynski2025mixed}. Instead, it is different from other nonequilibrium FRRs presented in the past~\cite{agarwal1972fluctuation,baiesi2009fluctuations,seifert2010fluctuation,prost2009generalized,altaner2016fluctuation,chun2023trade,shiraishi2023introduction,gao2024thermodynamic,tesser2024out,maes2020response}. In particular, it differs from the result from Refs.~\cite{agarwal1972fluctuation,baiesi2009fluctuations,seifert2010fluctuation}, which enables one to express a response of any trajectory observable for Markov jump processes (including state observable) in terms of its covariance with a specially constructed auxiliary observable that is not a state observable.

\subsection{Upper bound for fluctuations of state observables}
We now show that our result further produces upper bounds for the variance of state observables $\langle\!\langle \mathcal{O} \rangle\!\rangle \equiv\langle\!\langle \mathcal{O},\mathcal{O} \rangle\!\rangle$ expressed in terms of the dynamic and thermodynamic quantities. To derive them, we use the result of Ref.~\cite{gabriela2023topologically}, bounding the response to symmetric and antisymmetric perturbations as 
\begin{align} \label{eq:bounds_gabriela}
\big|d_{B_e} \mathcal{O} \big| \leq [\![ \mathcal{O} ]\!]\tanh{(\mathcal{F}_\text{max}/4)}\,,\quad
    \big|d_{S_e} \mathcal{O} \big| \leq [\![ \mathcal{O} ]\!] \,, 
\end{align}
where $[\![ \mathcal{O} ]\!] \equiv (\max_n o_n - \mathcal{O})(\mathcal{O}-\min_n o_n)/(\max_n o_n-\min_n o_n)$ and $\mathcal{F}_\text{max}=\text{max}_c|\mathcal{F}_c|$ is the maximum of the cycle affinities $\mathcal{F}_c$; see Ref.~\cite{gabriela2023topologically} for even tighter bounds dependent on the network topology. Inserting \cref{eq:bounds_gabriela} into \cref{eq:covar-exact}, we obtain the mentioned bounds for the variances,
\begin{subequations} \label{eq:bound-upper-var}
\begin{align}
   \langle\!\langle \mathcal{O} \rangle\!\rangle &\leq [\![ \mathcal{O} ]\!]^2 \tanh^2{(\mathcal{F}_\text{max}/4)}\sum_{e}\frac{\tau_e}{j_e^2} \,,\\
    \langle\!\langle \mathcal{O} \rangle\!\rangle &\leq  [\![ \mathcal{O} ]\!]^2 \sum_{e}\frac{4}{\tau_e} \leq  \frac{4|\mathcal{E}| \times [\![ \mathcal{O} ]\!]^2}{\min_e \tau_e} \,,
\end{align}
\end{subequations}
where $|\mathcal{E}|$ is the number of edges. We notice that the second inequality can be used to upper bound the minimum edge traffic, $\min_e \tau_e \leq 4|\mathcal{E}| \times [\![ \mathcal{O} ]\!]^2/ \langle\!\langle \mathcal{O} \rangle\!\rangle$, provided the network topology is known. Notably, the inequalities \eqref{eq:bound-upper-var} appear to be the first known upper bounds for fluctuations of state observables. We further emphasize that these inequalities are expressed solely in terms of the dynamic and thermodynamic quantities of the system, in contrast to the previously derived upper bound for fluctuations of jump observables~\cite{bakewell2023general} that included also a spectral gap of the rate matrix $\mathbb{W}$.

\subsection{Relation to fluctuation-response inequalities}
Let us now discuss the relation of our FRRs to previous results. First, they are related to fluctuation-response inequalities, derived in Refs.~\cite{kwon2024fluctuation,zheng2024information, liu2024dynamical} for the time-dependent variance $\langle\!\langle \mathcal{O}(t) \rangle\!\rangle \equiv\langle\!\langle \mathcal{O}(t),\mathcal{O}(t) \rangle\!\rangle$,
\begin{subequations} \label{eq:var-ineq}
\begin{align}
\label{eq:var-ineq-gen}
\langle\!\langle\mathcal{O}(t) \rangle\!\rangle  &\geq \sum_{\pm e}\frac{1}{\varphi_{\pm e}}[d_{X_{\pm e}}\mathcal{O}(t)]^2\ \\
\label{eq:var-ineq-sym}
  &= \sum_{e}\frac{\tau_e}{j_e^2}[d_{B_e}\mathcal{O}(t)]^2\ \\
\label{eq:var-ineq-antisym}
 & = \sum_{e} \frac{4}{\tau_e} [d_{S_e}\mathcal{O}(t)]^2\,,
\end{align}
\end{subequations}
where ``='' sign denotes that the terms at the right-hand side are equal to each other (see Appendix~\ref{app:proof-identities-var}). In that context, $d_{p} \mathcal{O}(t)$ is the dynamic response to the perturbations introduced at $t=0$, for a system initialized in the stationary state of an unperturbed rate matrix. In the long-time limit $t \rightarrow \infty$, it converges to the static response $d_p \mathcal{O}$ considered in our paper. Notably, our result~\eqref{eq:covar-exact} proves a numerical conjecture from Ref.~\cite{kwon2024fluctuation} that the above inequalities (and thus the version of the Cram\'{e}r-Rao bound used to derive them) saturate for $t \rightarrow \infty$, which may have further implications for understanding the long-time behavior of Markov jump processes.

The above relations have also been used to derive the so called Response TURs and KURs, which bound the response to kinetic or thermodynamic perturbations by the stationary entropy production rate $\dot{\sigma}=\sum_e j_e \ln(\varphi_{+e}/\varphi_{-e})$ or the total traffic $\mathcal{T}=\sum_e \tau_e$. To that end, one introduces single parameters $\kappa$, $\varepsilon$ and $\eta$ that parametrize the edge parameters as $X_{\pm e}=X_{\pm e}(\kappa)$, $B_e=B_e(\varepsilon)$ and $S_e=S_e(\eta)$. Applying the Cauchy-Schwarz inequality, one obtains~\cite{kwon2024fluctuation, zheng2024information, liu2024dynamical}
\begin{subequations} \label{eq:tur}
\begin{align}
\label{eq:tur-gen}\frac{[d_{\kappa}\mathcal{O}(t)]^2}{\langle\!\langle\mathcal{O}(t) \rangle\!\rangle }  &\leq \max_{\pm e} |d_\kappa X_{\pm e}|^2 \mathcal{T} \,, \\
\label{eq:tur-sym}\frac{[d_{\varepsilon}\mathcal{O}(t)]^2}{\langle\!\langle\mathcal{O}(t) \rangle\!\rangle }  &\leq \max_e |d_\varepsilon B_e|^2 \min (\dot{\sigma}/2,\mathcal{T}) \,, \\
\label{eq:tur-antisym}
\frac{[d_{\eta}\mathcal{O}(t)]^2}{\langle\!\langle\mathcal{O}(t) \rangle\!\rangle }  &\leq \frac{\max_e |d_\eta S_e|^2\mathcal{T}}{4} \,.
\end{align}
\end{subequations}

\subsection{Response to vertex perturbations}
Here we complement Eqs.~\eqref{eq:var-ineq}--\eqref{eq:tur} by deriving (see Appendix~\ref{app:vertex-bounds}) similar inequalities, bounding response to perturbations of vertex parameters $V_n$. The first of them reads
\begin{align} \label{eq:vertex-ineq-dyn}
\langle\!\langle \mathcal{O}(t) \rangle\!\rangle \geq \sum_{n} \frac{[d_{V_{n}} \mathcal{O}(t)]^2}{\pi_n |W_{nn}|} \,.
\end{align}
In particular, the precision of the response to perturbation associated with each vertex $n$ is bounded by the probability of that state $\pi_n$ and the escape rate from that state $|W_{nn}|$:
\begin{align} \label{eq:vertex-ineq-dyn-sing}
\forall_n: \quad \frac{[d_{V_n} \mathcal O(t)]^2}{\langle\!\langle \mathcal{O}(t) \rangle\!\rangle} \leq \pi_n |W_{nn}| \,.
\end{align}
Parametrizing the vertex parameters using a single control parameter $\nu$, $V_n=V_n(\nu)$, we further get
\begin{align}
\label{eq:tur-vert}
\frac{[d_{\nu}\mathcal{O}(t)]^2}{\langle\!\langle\mathcal{O}(t) \rangle\!\rangle }  &\leq \max_n |d_\nu V_n|^2 \mathcal{T} \,.
\end{align}
These results have a particularly simple interpretation for the static response ($t \rightarrow \infty$) at thermodynamic equilibrium $\pi_n \propto e^{-\beta E_n}$, where $E_n$ are the state energies and $\beta=1/(k_B T)$ is the inverse temperature. Then, even though the vertex parameters cannot be unequivocally identified with state energies [see the discussion below \cref{eq:rates-model}], the static responses to vertex perturbations are directly proportional to the responses to perturbations of state energies: $d_{V_n} \mathcal{O}=\beta^{-1} d_{E_n} \mathcal{O}$ (see Appendix~\ref{app:vertex-energy}). Consequently, Eqs.~\eqref{eq:vertex-ineq-dyn}--\eqref{eq:tur-vert} bound the precision of equilibrium response to energy perturbations.

\subsection{Relation to ``occupation uncertainty relation''}
We now note that the long-time version of Eq.~\eqref{eq:vertex-ineq-dyn} is equivalent to another bound on $\langle\!\langle\mathcal{O}\rangle\!\rangle$, nicknamed ``occupation uncertainty relation'', recently derived in Ref.~\cite{macieszczak2024occupation}. To show that, we combine Eq.~\eqref{eq:vertex-ineq-dyn} with the exact relation for the static response to vertex perturbation, $d_{V_n} \pi_m=\pi_n (\pi_m-\delta_{nm})$~\cite{owen2020universal,aslyamov2024nonequilibrium}. In this way, we reproduce the result of Ref.~\cite{macieszczak2024occupation} (see Appendix~\ref{app:proof-equivalence-macieszczak}),
\begin{align} \label{eq:bound-macieszczak}
\langle\!\langle \mathcal{O} \rangle\!\rangle = \boldsymbol{o}^\intercal \mathbb{C} \boldsymbol{o} \geq \boldsymbol{o}^\intercal \mathbb{Q} \cdot \mathbb{\Lambda}^{-1} \cdot \text{diag}(\boldsymbol{\pi}) \cdot \mathbb{Q}^\intercal \boldsymbol{o} \,,
\end{align}
where $(\mathbb{Q})_{mn} \equiv \delta_{nm}-\pi_m$ and $\mathbb{\Lambda} \equiv \text{diag}(\ldots,|W_{nn}|,\ldots)$. 
In this way, the response theory framework shows a close connection between previously unrelated results, Eqs.~\eqref{eq:var-ineq} and \cref{eq:bound-macieszczak}. We note that the above inequality complements our upper bounds~\eqref{eq:bound-upper-var}. Like them, it is expressed explicitly in terms of transition rates $W_{\pm e}$ and state probabilities $\pi_n$, rather than objects such as the Drazin inverse $\mathbb{W}^D$ or the spectral gap of $\mathbb{W}$. This may facilitate the analytic study of fluctuations, as the latter objects are often difficult to handle analytically.

\section{Example}
\begin{figure}
    \centering
    \includegraphics[width=\linewidth]{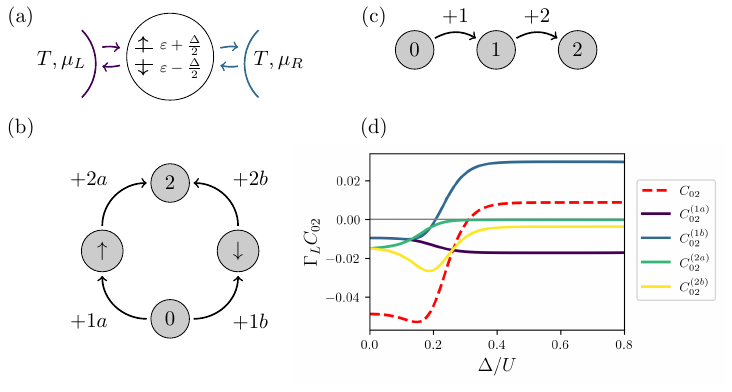}
    \caption{(a) Scheme of a quantum dot with two spin levels whose energies are Zeeman-splitted by the magnetic field. $\varepsilon$ denotes the average level energy, and $\Delta$ denotes the Zeeman splitting. The dot is coupled to two reservoirs $L$ and $R$ with chemical potentials $\mu_L=V/2$, $\mu_R=-V/2$, and temperature $T$. (b) The system dynamics is described by four-state Markov network with an empty state $0$, states occupied by single electron with spin $\uparrow$ or $\downarrow$, and doubly occupied state $2$. The transition rates read as $W_{\pm e}=\sum_{r \in \{L,R \}} \Gamma_r f[(E_{t(\pm e)}-E_{s(\pm e)} \mp \mu_r)/k_B T]$, where $\Gamma_r$ are tunnel couplings to reservoirs, $f(x) \equiv 1/[1+\exp(x)]$ is the Fermi-Dirac distribution, and the state energies read $E_0=0$, $E_\uparrow=\varepsilon+\Delta/2$, $E_\downarrow=\varepsilon-\Delta/2$, $E_2=U+2\varepsilon$, $U \geq 0$ being the Coulomb coupling. (c) For $\Delta=0$, the electron-tunneling is spin independent: $W_{\pm 1a}=W_{\pm 1b}$, $W_{\pm 2a}=W_{\pm 2b}$. As a result, the system dynamics can be described using coarse-grained one-dimensional Markov network, where $1$ is the union of states $\uparrow$ and $\downarrow$, $W_{+1}=2W_{+1a}$, $W_{-1}=W_{-1a}$, $W_{+2}=W_{+2a}$, $W_{-2}=2W_{-2a}$. (d) The covariance $C_{02}$ and its components $C_{02}^{(e)} \equiv 4 d_{S_e} \pi_0 d_{S_e} \pi_2/\tau_e$ as a function of $\Delta$. Parameters: $\Gamma_R=0.2\Gamma_L$, $k_B T=0.02U$, $\varepsilon=-0.3U$, $V=1.6U$. The analytic expressions for $C_{02}^{(e)}$ are presented in Appendix~\ref{app:c02e}.}
    \label{fig:fig-QD}
\end{figure}

Finally, let us illustrate how our identities allow one to understand the mechanistic origin of fluctuations, and thus connect their behavior to the topology of the Markov network. To that end, we employ the quantum dot model shown in Fig.~\ref{fig:fig-QD}, which has been previously theoretically studied in Ref.~\cite{stegmann2017violation}. We note that in such systems the charge states can be monitored in real time using charge counting techniques~\cite{gustavsson2006counting,gustavsson2009electron}, which could enable the experimental verification of our results. The notable feature of our example is that its effective topology depends on the Zeeman splitting $\Delta$. For $\Delta=0$, it can be effectively described using a coarse-grained one-dimensional model~[Fig.~\ref{fig:fig-QD}~(c)]. This is also approximately true for sufficiently small $\Delta$, when the tunneling is nearly spin-independent: for all $e$, $W_{+ e} \approx \Gamma_L$, $W_{-e} \approx \Gamma_R$. For the parameters considered, this occurs for $\Delta \lessapprox 0.2 U$. In such a case, even without explicit calculations, using FRR~\eqref{eq:covmat-exact-antisym} one may predict that covariance $C_{02}$ must be negative. This is because by enhancing the parameters $S_1$ and $S_2$ one reduces the probability $\pi_0$ and increases $\pi_2$, so that the terms $d_{S_e} \pi_0 d_{S_e} \pi_2$ are negative. In fact, for the model in Fig.~\ref{fig:fig-QD}~(c), the state probabilities can be calculated using an analytic formula
\begin{align} \label{eq:prob-onedim}
\pi_n=\pi_0 \prod_{e=1}^n \exp(S_e) \,,
\end{align}
where $S_e=\ln(W_{+e}/W_{-e})$, and $\pi_0$ is determined by the normalization condition $\sum_{n=0}^N \pi_n=1$, where $N=2$ is the number of states. As a result, one finds 
\begin{align} \label{eq:resp-onedim}
d_{S_e} \pi_n=\pi_n \left[\Theta(n-e) -\sum_{k=e}^N \pi_k\right] \,,
\end{align} where $\Theta(x)$ is the Heaviside theta with $\Theta(0)=1$. This implies that $d_{S_e} \pi_0<0$ and $d_{S_e} \pi_2 >0$. We note here that Eqs.~\eqref{eq:prob-onedim}-\eqref{eq:resp-onedim}, which allow determining the state responses (and thus, using FRRs, also the fluctuations), can be applied to any one-dimensional Markov model (with arbitrary $N$). Such models (so called ``birth-and-death'' processes) are used in many contexts, including chemical bistability (Schl{\"o}gl model)~\cite{schlogl1972chemical,vellela2009stochastic,ge2009thermodynamic}, bistable electric circuits~\cite{landauer1962fluctuations,hanggi1982stochastic}, lasers~\cite{carmichael1999statistical1}, magnetic systems (Curie-Weiss model)~\cite{herpich2020njp,meibohm2022finite,meibohm2023landau,ptaszynski2024critical}, coupled heat engines~\cite{vroylandt2018collective,vroylandt2020efficiency}, population genetics~\cite{novozhilov2006biological}, or disease spread~\cite{nieddu2022characterizing}.

On the other hand, we observe that the covariance $C_{02}$ becomes positive for $\Delta \gtrapprox 0.3U$ [Fig.~\ref{fig:fig-QD}~(d)]. Based on our discussion, this requires that the effective topology of the system has changed, so that it can no longer be described using the one-dimensional model, but one needs a full cyclic model [Fig.~\ref{fig:fig-QD}~(b)]. This happens because the transition rates become strongly spin dependent, with $W_{+2a} \approx \Gamma_L$, $W_{-2a} \approx \Gamma_R$ while $W_{+2b} \approx 0$, $W_{-2b} \approx \Gamma_L+\Gamma_R$. This shows that analysis of fluctuations of state observables, combined with qualitative and analytic insight provided by our FRRs, can help to infer the topology of the underlying Markov process. We also note that FRR~\eqref{eq:covmat-exact-antisym} allows one to decompose the covariance $C_{02}$ into a sum of individual components $C_{02}^{(e)} \equiv 4 d_{S_e} \pi_0 d_{S_e} \pi_2/\tau_e$. This enables a mechanistic interpretation of the behavior of fluctuations in terms of the response properties of the system. To illustrate this, in Fig.~\ref{fig:fig-QD}~(d) we show that $C_{02}$ becomes positive because the term $C_{02}^{(1b)}$ becomes positive. This occurs because the transition $W_{+2b}$ becomes suppressed, so increasing $S_{1b}$ increases $\pi_\downarrow$ while reducing both $\pi_0$ and $\pi_2$.

\section{Concluding remarks}
As illustrated by our example, the FRRs~\eqref{eq:covar-exact} sometimes allow one to predict signs of covariances of state observables based only on the topology of the Markov network. This may help to infer the network topology based on measured data, e.g., by providing input into physics-informed machine learning protocols. We hope this will inspire further research on how topology universally governs the properties of responses and, consequently, fluctuations. On the methodological side, as illustrated by derivation of FRRs in Appendix~\ref{app:proof-frr}, our article demonstrates that the recently developed algebraic approach to responses~\cite{aslyamov2024nonequilibrium,aslyamov2024general,aslyamov2024frr} and fluctuations~\cite{lapolla2018unfolding,lapolla2019manifestations,lapolla2020spectral,landi2023current} can be a powerful tool to determine the universal properties of these quantities. Examining the tightness of the bounds~\eqref{eq:bound-upper-var} and~\eqref{eq:vertex-ineq-dyn}--\eqref{eq:tur-vert} for physically relevant setups is an interesting perspective. Future studies may also be concerned with the generalization of FRRs to continuous-space Langevin dynamics~\cite{gao2022thermodynamic,gao2024thermodynamic}, where state observables have received significant interest~\cite{levy1940certains,dhar1999residence,godreche2001statistics,majumdar2002local,majumdar2007brownian,lapolla2018unfolding,lapolla2019manifestations,lapolla2020spectral,singh2019generalised}.

\section*{Author's note}
We note that Ref.~\cite{bao2024nonequilibrium} used our results to generalize FRRs~\eqref{eq:covar-exact} to nonlinear responses to perturbations of transition rates.

\begin{acknowledgments}
K.P., T.A. and M.E. acknowledge the financial support from, respectively, project No.\ 2023/51/D/ST3/01203 funded by the National Science Centre, Poland, 
project ThermoElectroChem (C23/MS/18060819) from Fonds National de la Recherche-FNR, Luxembourg,  
project TheCirco (INTER/FNRS/20/15074473) funded by FRS-FNRS (Belgium) and FNR (Luxembourg).

\end{acknowledgments}

\appendix

\section{Equivalence of Eq.~\eqref{eq:covmat-expr} and Refs.~\cite{lapolla2018unfolding,lapolla2019manifestations,lapolla2020spectral}} \label{app:equiv-covmat}
To demonstrate that Eq.~\eqref{eq:covmat-expr} is equivalent to the result of Refs.~\cite{lapolla2018unfolding,lapolla2019manifestations,lapolla2020spectral}, we express the latter using our notation,
\begin{align} \nonumber
C_{mn}&=-\sum_{k} \frac{1}{\lambda_k} \left( \boldsymbol{1}^\intercal \mathds{1}^{(m)} \boldsymbol{r}_k \boldsymbol{l}_k^\intercal \mathds{1}^{(n)} \boldsymbol{\pi}+\boldsymbol{1}^\intercal \mathds{1}^{(n)} \boldsymbol{r}_k \boldsymbol{l}_k^\intercal \mathds{1}^{(m)} \boldsymbol{\pi} \right) \\
&=-\boldsymbol{1}^\intercal \mathds{1}^{(m)} \mathbb{W}^D \mathds{1}^{(n)} \boldsymbol{\pi}-\boldsymbol{1}^\intercal \mathds{1}^{(n)} \mathbb{W}^D \mathds{1}^{(m)} \boldsymbol{\pi} \,,
\end{align}
where $\boldsymbol{1} \equiv (1,1,\ldots)^\intercal$, $\mathds{1}^{(n)}=\text{diag}(\delta_{1n},\delta_{2n},\ldots)$, $\lambda_k$ are nonzero eigenvalues of $\mathbb{W}$, and $\boldsymbol{l}_k^\intercal$, $\boldsymbol{r}_k$ are the associated left and right eigenvectors, $\boldsymbol{l}_k^\intercal \mathbb{W}=\lambda_k \boldsymbol{l}_k^\intercal$, $\mathbb{W} \boldsymbol{r}_k=\lambda_k \boldsymbol{r}_k$. In the second equality, we use the spectral decomposition of the Drazin inverse, $\mathbb{W}^D=\sum_k \boldsymbol{r}_k \boldsymbol{l}_k^\intercal/\lambda_k$~\cite{landi2023current}. Consequently,
\begin{align}
C_{mn}=-(\mathbb{W}^D)_{mn} \pi_n-(\mathbb{W}^D)_{nm} \pi_m \,,
\end{align}
which is equivalent to Eq.~\eqref{eq:covmat-expr}.

\section{Derivation of Eq.~\eqref{eq:response-drazin}} \label{app:der-response}
Equation~\eqref{eq:response-drazin} can be derived by acting on Eq.~\eqref{eq:NESS} with the derivative $d_{p}$. This yields
\begin{align}
\mathbb{W}d_{p} \boldsymbol{\pi} =-(d_{p} \mathbb W) \boldsymbol{\pi} \,.
\end{align}
Acting on both sides of the above expression with the Drazin inverse $\mathbb{W}^D$ we get
\begin{align}
\mathbb{W}^D \mathbb{W}d_{p} \boldsymbol{\pi} =d_{p} \boldsymbol{\pi}-\boldsymbol{\pi} (\boldsymbol{1}^\intercal d_{p} \boldsymbol{\pi})=-\mathbb{W}^D (d_{p} \mathbb W) \boldsymbol{\pi} \,.
\end{align}
where we used $\mathbb{W}^D \mathbb{W}=\mathds{1}-\boldsymbol{\pi} \boldsymbol{1}^\intercal$~\cite{landi2023current}. Since the response vector $d_{p} \boldsymbol{\pi}$ is traceless, $\boldsymbol{1}^\intercal d_{p} \boldsymbol{\pi}=0$, which yields Eq.~\eqref{eq:response-drazin}. 

\section{Proof of fluctuation-response relations~\eqref{eq:covmat-exact}} \label{app:proof-frr}

Here we prove our main result. Let us first explicitly express the responses of state probabilities included in the right-hand side of Eqs.~\eqref{eq:covmat-exact-gen}--\eqref{eq:covmat-exact-antisym} in terms of Eq.~\eqref{eq:response-drazin}. For generic perturbations, we get
\begin{align}
d_{X_{\pm e}}\mathbb{W}=
\begin{blockarray}{ccccccccc}
& &  \color{gray} \dots & \color{gray} s(\pm e) & \color{gray}\dots & \\
\begin{block}{cc (ccccccc)}
\color{gray} \vdots & &  \phantom{0} & \phantom{0} & \phantom{0} & \phantom{0} \\
\color{gray} t(\pm e) & & \phantom{0} & W_{ \pm e} & \phantom{0} & \phantom{0} \\
\color{gray} \vdots & &  \phantom{0} & \phantom{0} & \phantom{0} & \phantom{0} \\
\color{gray} s(\pm e) & & \phantom{0} & -W_{\pm e} & \phantom{0} & \phantom{0} \\
\color{gray} \vdots & &  \phantom{0} & \phantom{0} & \phantom{0} & \phantom{0}\\
\end{block}
\end{blockarray}\,\,,
\end{align}
and thus $d_{X_{\pm e}} \pi_m =\varphi_{\pm e} [(\mathbb{W}^D)_{m s(\pm e)} -(\mathbb{W}^D)_{m t(\pm e)}]$ where $\varphi_{\pm e} \equiv W_{\pm e} \pi_{s (\pm e)}$. Consequently, using $t(\pm e)=s(\mp e)$, we have $d_{B_e} \pi_m=d_{X_{+e}} \pi_m+d_{X_{-e}} \pi_m=j_e[(\mathbb{W}^D)_{m s(+e)} -(\mathbb{W}^D)_{m t(+e)}]$ and $d_{S_e} \pi_m=(d_{X_{+e}} \pi_m-d_{X_{-e}} \pi_m)/2=\tau_e[(\mathbb{W}^D)_{m s(+e)}-(\mathbb{W}^D)_{m t(+e)}]/2$. As a result, the right-hand sides of Eqs.~\eqref{eq:covmat-exact-gen}--\eqref{eq:covmat-exact-antisym} are the same and can be expanded as
\begin{align} \nonumber
&\sum_{\pm e}\frac{1}{\varphi_{\pm e}}d_{X_{\pm e}} \pi_m d_{X_{\pm e}} \pi_n \\ \nonumber
&=\sum_{e}\frac{\tau_e}{j_e^2}d_{B_e} \pi_m d_{B_e} \pi_n = \sum_{e} \frac{4}{\tau_e} d_{S_e} \pi_m d_{S_e} \pi_n  \\ \nonumber &=\sum_e \tau_e [(\mathbb{W}^D)_{m s(+e)}-(\mathbb{W}^D)_{m t(+e)}] \\ \label{eq:expansion-rhs-frr} &\times [(\mathbb{W}^D)_{n s(+e)}-(\mathbb{W}^D)_{n t(+e)}] \,.
\end{align}
We now replace the summation over the edges $e$ with the summation over the pairs of states $k$ and $l$. To that end, we define the sum of all traffics at the edges connecting states $k$ and $l$, $\tau_{kl} \equiv \sum_e [\delta_{ks(+e)}\delta_{lt(+e)}+\delta_{kt(+e)}\delta_{ls(+e)}]\tau_e=W_{kl} \pi_l+W_{lk} \pi_k$. The above expression then becomes equal to
\begin{align} \nonumber
&\sum_{k,l<k} \tau_{kl}  (\mathbb{W}^D)_{m k} [(\mathbb{W}^D)_{n k}-(\mathbb{W}^D)_{nl}]  \\ \nonumber
&-\sum_{k,l<k} \tau_{kl} (\mathbb{W}^D)_{m l} [(\mathbb{W}^D)_{n k}-(\mathbb{W}^D)_{nl}] \\ \nonumber
&=\sum_{k,l<k} \tau_{kl}  (\mathbb{W}^D)_{m k} [(\mathbb{W}^D)_{n k}-(\mathbb{W}^D)_{nl}]  \\ \nonumber
&-\sum_{l,k<l} \tau_{lk} (\mathbb{W}^D)_{m k} [(\mathbb{W}^D)_{n l}-(\mathbb{W}^D)_{nk}]
\\ \label{eq:expansion-sums} &=\sum_{k,l \neq k} \tau_{kl}  (\mathbb{W}^D)_{m k} [(\mathbb{W}^D)_{n k}-(\mathbb{W}^D)_{nl}] \,,
\end{align}
where in the first step we flipped the indices $k$ and $l$ in the second sum, while in the second step we used $\tau_{lk}=\tau_{kl}$ and noted $\sum_{k,l<k} (\cdot)+\sum_{l,k<l} (\cdot)=\sum_{k,l<k} (\cdot)+\sum_{k,l>k} (\cdot)=\sum_{k,l \neq k} (\cdot)$.

We now show that the covariance matrix elements $C_{mn}$ are equal to the result of \cref{eq:expansion-sums}. First, we note
\begin{align}
\mathbb{W}^D \cdot \text{diag}(\boldsymbol{\pi})=\mathbb{W}^D \cdot \text{diag}(\boldsymbol{\pi}) \cdot 
(\mathbb{W}^D \mathbb{W} )^\intercal \,,
\end{align}
which results from the identities $(\mathbb{W}^D \mathbb{W} )^\intercal=\mathds{1}-\boldsymbol{1} \boldsymbol{\pi}^\intercal$ and $\mathbb{W}^D \cdot \text{diag}(\boldsymbol{\pi}) \cdot \boldsymbol{1} =\mathbb{W}^D \boldsymbol{\pi}=0$~\cite{landi2023current}. As a result
\begin{align} \nonumber
&[\mathbb{W}^D \cdot \text{diag}(\boldsymbol{\pi})]_{mn}=[\mathbb{W}^D \cdot \text{diag}(\boldsymbol{\pi}) \cdot \mathbb{W}^\intercal (\mathbb{W}^D)^\intercal]_{mn} \\&=\sum_{k,l} (\mathbb{W}^D)_{mk} \pi_k W_{lk} (\mathbb{W}^D)_{nl}\,. \end{align}
Analogously,
\begin{align} \nonumber
&[\mathbb{W}^D \cdot \text{diag}(\boldsymbol{\pi})]^\intercal_{mn}=[\mathbb{W}^D \cdot \text{diag}(\boldsymbol{\pi})]_{nm} \\&=\sum_{k,l} (\mathbb{W}^D)_{nk} \pi_k W_{lk} (\mathbb{W}^D)_{ml}=\sum_{k,l} (\mathbb{W}^D)_{mk} \pi_l W_{kl} (\mathbb{W}^D)_{nl}\,, \end{align}
where in the last step we flipped indices $k$ and $l$. Consequently,
\begin{align} \nonumber
&C_{mn}=-[\mathbb{W}^D \cdot \text{diag}(\boldsymbol{\pi})]_{mn}-[\mathbb{W}^D \cdot \text{diag}(\boldsymbol{\pi})]_{mn}^\intercal \\ \label{eq:cmn-der1}
&=-\sum_{k,l} ( W_{lk}\pi_k + W_{kl} \pi_l ) (\mathbb{W}^D)_{mk} (\mathbb{W}^D)_{nl} \,.
\end{align}
Let us now focus on the term $( W_{lk}\pi_k + W_{kl} \pi_l )$. For $k \neq l$, it is equal to $\tau_{kl}$ defined below \cref{eq:expansion-rhs-frr}. For $k=l$, we note that $-W_{kk} \pi_k=\sum_{l \neq k} W_{lk} \pi_k$ is the probability flux out of the state $k$. Due to Kirchhoff's law, it is equal to the flux to that state, $\sum_{l \neq k} W_{kl} \pi_l$. Consequently, $( W_{kk}\pi_k + W_{kk} \pi_k )=-\sum_{l \neq k} (W_{lk} \pi_k+W_{kl} \pi_l)=-\sum_{l \neq k} \tau_{kl}$. As a result, Eq.~\eqref{eq:cmn-der1} becomes
\begin{align}
&C_{mn}=\sum_{k,l \neq k} \tau_{kl}  (\mathbb{W}^D)_{m k} [(\mathbb{W}^D)_{n k}-(\mathbb{W}^D)_{nl}] \,,
\end{align}
which is equal to the result of Eq.~\eqref{eq:expansion-sums}. This concludes the proof.

\section{Proof of identities in \cref{eq:var-ineq}} \label{app:proof-identities-var}
Here, we prove that the right-hand sides of Eqs.~\eqref{eq:var-ineq-gen}--\eqref{eq:var-ineq-antisym} are identical. First, we note that the identity between the right-hand sides of Eqs.~\eqref{eq:var-ineq-sym} and \eqref{eq:var-ineq-antisym} has been proven in Ref.~\cite{kwon2024fluctuation}. To prove the identity between the right-hand sides of Eqs.~\eqref{eq:var-ineq-gen} and \eqref{eq:var-ineq-antisym} we use the total derivatives to express the responses as
$d_{B_e} \mathcal{O}(t)=d_{X_{+e}}\mathcal{O}(t)+d_{X_{-e}}\mathcal{O}(t) $ and $d_{S_e} \mathcal{O}(t)=[d_{X_{+e}}\mathcal{O}(t)-d_{X_{-e}}\mathcal{O}(t)]/2$. This yields $d_{X_{\pm e}}\mathcal{O}(t)=d_{B_e} \mathcal{O}(t)/2 \pm d_{S_e} \mathcal{O}(t)$. Using the identity $d_{B_e} \mathcal{O}(t)/2=(j_e/\tau_e) d_{S_e} \mathcal{O}(t)$ from Ref.~\cite{kwon2024fluctuation} we have
\begin{align}
d_{X_{\pm e}}\mathcal{O}(t)=\left( \frac{j_{e}}{\tau_e} \pm 1 \right) d_{S_e} \mathcal{O}(t)=\pm \frac{2 \varphi_{\pm e}}{\tau_e} d_{S_e} \mathcal{O}(t) \,,
\end{align}
where in the second step we used $j_e \equiv \varphi_{+e} -\varphi_{-e}$ and $\tau_e \equiv \varphi_{+e} +\varphi_{-e}$. Inserting this into the right-hand side of Eq.~\eqref{eq:var-ineq-gen}, one obtains the right-hand side of Eq.~\eqref{eq:var-ineq-antisym}.

\section{Derivation of Eqs.~\eqref{eq:vertex-ineq-dyn} and~\eqref{eq:tur-vert}} \label{app:vertex-bounds}
To derive Eq.~\eqref{eq:vertex-ineq-dyn}, we rewrite Eq.~\eqref{eq:var-ineq-gen} as
\begin{align} \nonumber \label{eq:bound-vertex-der}
&\langle\!\langle \mathcal{O}(t) \rangle\!\rangle \geq \sum_{n} \sum_{\pm e} \frac{[\delta_{ns(\pm e)}  d_{X_{\pm e}} \mathcal{O}(t)]^2}{W_{\pm e} \pi_{s(\pm e)}} \\ &\geq  \sum_{n} \frac{[\sum_{\pm e} \delta_{ns(\pm e)}  d_{X_{\pm e}} \mathcal{O}(t)]^2}{\sum_{\pm e} \delta_{ns(\pm e)} W_{\pm e} \pi_{s(\pm e)}}  \,.
\end{align}
Here, in the second step, we applied Sedrakyan's inequality, $\sum_i a_i^2/b_i \geq (\sum_i a_i)^2/(\sum_i b_i)$, which holds for real $a_i$ and positive $b_i$, to each sum with a given $n$. We then notice
$d_{V_n} \mathcal{O}=\sum_{\pm e} \delta_{ns(\pm e)} d_{X_{\pm e}} \mathcal{O}$ and $\sum_{\pm e} \delta_{ns(\pm e)} W_{\pm e} \pi_{s(\pm e)}=-W_{nn} \pi_n$. As a result, we obtain \cref{eq:vertex-ineq-dyn}. The bound~\eqref{eq:tur-vert} is then obtained by applying Sedrakyan's inequality to \cref{eq:vertex-ineq-dyn}.

\section{Proof of relation $d_{V_n} \mathcal{O}=\beta^{-1} d_{E_n} \mathcal{O}$} \label{app:vertex-energy}
At thermodynamic equilibrium, the state probabilities are expressed as $\pi_n=e^{-\beta E_n}/Z$, where $Z=\sum_m e^{-\beta E_m}$ is the partition function. This yields $d_{E_n} \pi_m=\beta \pi_n(\pi_m-\delta_{nm})$. At the same time, for a generic Markov network, $d_{V_n} \pi_m=\pi_n(\pi_m-\delta_{nm})$~\cite{owen2020universal, aslyamov2024nonequilibrium}. Thus, at equilibrium, $d_{V_n} \mathcal{O}=\beta^{-1} d_{E_n} \mathcal{O}$.

\section{Derivation of Eq.~\eqref{eq:bound-macieszczak}}\label{app:proof-equivalence-macieszczak}
To derive Eq.~\eqref{eq:bound-macieszczak}, we use the relation $d_{V_n} \pi_m=\pi_n(\pi_m-\delta_{nm})$ to expand Eq.~\eqref{eq:vertex-ineq-dyn} (for $t \rightarrow \infty$) as
\begin{align} \nonumber
&\langle\!\langle \mathcal{O} \rangle\!\rangle \geq \sum_{n} \frac{[d_{V_{n}} \mathcal{O}]^2}{\pi_n |W_{nn}|}=\sum_n \frac{(\boldsymbol{o}^\intercal d_{V_n} \boldsymbol{\pi})^2}{\pi_n |W_{nn}|} \\  &\nonumber =\sum_n \frac{[\sum_m o_m \pi_n (\pi_m-\delta_{nm})]^2}{\pi_n |W_{nn}|}=\sum_{n} \frac{\pi_n [\sum_m o_m (\pi_m-\delta_{nm})]^2}{ |W_{nn}|} \\  \label{eq:proof-equivalence-macieszczak} &=\boldsymbol{o}^\intercal \mathbb{Q} \cdot \mathbb{\Lambda}^{-1} \cdot \text{diag}(\boldsymbol{\pi}) \cdot \mathbb{Q}^\intercal \boldsymbol{o} \,,
\end{align}
which gives Eq.~\eqref{eq:bound-macieszczak}; here, in the last step we used $[\boldsymbol{o}^\intercal \mathbb{Q}]_n=[ \mathbb{Q}^\intercal \boldsymbol{o}]_n=\sum_m o_m (\delta_{nm}-\pi_m)$.

\section{Analytic expressions for $C_{02}^{(e)}$} \label{app:c02e}
To determine the terms $
C_{02}^{(e)} \equiv 4 d_{S_e} \pi_0 d_{S_e} \pi_2/\tau_e$, we first calculate the state probabilities as $
\pi_n=a_n/({\sum_m a_m})$, where
\begin{align} \nonumber
a_0 &= W_{-1a} W_{-2a} (W_{-1b}+W_{+2b})+W_{-1b} W_{-2b} (W_{-1a}+W_{+2a}) \,, \\
\nonumber a_{\uparrow} &=W_{+1a} W_{-1b} (W_{-2a}+W_{-2b})+W_{-2a} W_{+2b} (W_{+1a} +W_{+1b}) \,, \\
\nonumber a_{\downarrow} &=W_{-1a} W_{+1b} (W_{-2a}+W_{-2b})+W_{+2a} W_{-2b} (W_{+1a} +W_{+1b}) \,, \\
a_2 &=W_{+1a} W_{+2a} (W_{-1b}+W_{+2b})+W_{+1b} W_{+2b}( W_{-1a}+W_{+2a}) \,.
\end{align}
The responses $d_{S_e} \pi_n$ are then given by the chain rule~\cite{owen2020universal}
\begin{align}
d_{S_e} \pi_n=\frac{1}{2} \left (W_{+e} \frac{d \pi_n}{dW_{+e}}-W_{-e} \frac{d \pi_n}{dW_{-e}} \right) \,.
\end{align}
Finally, the edge traffics can be calculated as $\tau_{1a}=W_{+1a} \pi_0+W_{-1a} \pi_\uparrow$, $\tau_{1b}=W_{+1b} \pi_0+W_{-1b} \pi_\downarrow$, $
\tau_{2a}=W_{+2a} \pi_\uparrow+W_{-2a} \pi_2$, $\tau_{2b}=W_{+2b} \pi_\downarrow+W_{-2b} \pi_2$.

\bibliography{biblio}

@article{maes2020response,
  title={Response theory: a trajectory-based approach},
  author={Maes, Christian},
  journal={Frontiers in Physics},
  volume={8},
  pages={229},
  year={2020},
  publisher={Frontiers Media SA},
doi={10.3389/fphy.2020.00229}
}

@article{ptaszynski2024dissipation,
  title={Dissipation bounds precision of current response to kinetic perturbations},
  author={Ptaszy{\'n}ski, Krzysztof and Aslyamov, Timur and Esposito, Massimiliano},
  journal={Phys. Rev. Lett.},
  volume={133},
  number={22},
  pages={227101},
  year={2024},
  publisher={APS},
doi = {10.1103/PhysRevLett.133.227101}
}

@article{van2024fundamental,
  title = {Fundamental Bounds on Precision and Response for Quantum Trajectory Observables},
  author = {Van Vu, Tan},
  journal = {PRX Quantum},
  volume = {6},
  issue = {1},
  pages = {010343},
  numpages = {22},
  year = {2025},
  month = {Mar},
  publisher = {American Physical Society},
  doi = {10.1103/PRXQuantum.6.010343},
  url = {https://link.aps.org/doi/10.1103/PRXQuantum.6.010343}
}

@article{mora2010limits,
  title={Limits of sensing temporal concentration changes by single cells},
  author={Mora, Thierry and Wingreen, Ned S},
  journal={Physical Review Letters},
  volume={104},
  number={24},
  pages={248101},
  year={2010},
  publisher={APS},
doi={10.1103/PhysRevLett.104.248101}
}

@article{harvey2023universal,
  title={Universal energy-accuracy tradeoffs in nonequilibrium cellular sensing},
  author={Harvey, Sarah E and Lahiri, Subhaneil and Ganguli, Surya},
  journal={Physical Review E},
  volume={108},
  number={1},
  pages={014403},
  year={2023},
  publisher={APS},
doi={10.1103/PhysRevE.108.014403}
}

@article{liu2024dynamical,
  title={Dynamical activity universally bounds precision of response in Markovian nonequilibrium systems},
  author={Liu, Kangqiao and Gu, Jie},
  journal={Communications Physics},
  volume={8},
  number={1},
  pages={1--9},
  year={2025},
  publisher={Nature Publishing Group},
doi={10.1038/s42005-025-01982-w}
}

@article{kwon2024fluctuation,
  title = {Fluctuation-Response Inequalities for Kinetic and Entropic Perturbations},
  author = {Kwon, Euijoon and Chun, Hyun-Myung and Park, Hyunggyu and Lee, Jae Sung},
  journal = {Phys. Rev. Lett.},
  volume = {135},
  issue = {9},
  pages = {097101},
  numpages = {8},
  year = {2025},
  month = {Aug},
  publisher = {American Physical Society},
  doi = {10.1103/h45s-4118},
  url = {https://link.aps.org/doi/10.1103/h45s-4118}
}

@book{stratonovich2012nonlinear,
  title={Nonlinear nonequilibrium thermodynamics I: linear and nonlinear fluctuation-dissipation theorems},
  author={Stratonovich, Rouslan L},
  volume={57},
  year={2012},
  publisher={Springer Science \& Business Media}
}

@article{shiraishi2023introduction,
  title={An Introduction to Stochastic Thermodynamics},
  author={Shiraishi, Naoto},
  journal={Fundamental Theories of Physics. Springer, Singapore},
  year={2023},
  publisher={Springer}
}

@book{kubo2012statistical,
  title={Statistical physics II: nonequilibrium statistical mechanics},
  author={Kubo, Ryogo and Toda, Morikazu and Hashitsume, Natsuki},
  volume={31},
  year={2012},
  publisher={Springer Science \& Business Media}
}

@article{dechant2019arxiv,
        title={Fluctuation-response inequality out of equilibrium}, 
      author={Andreas Dechant and Shin-ichi Sasa},
      year={2019},
  journal={arXiv preprint},
  year={2018},
 url={https://arxiv.org/abs/1804.08250}
}

@article{santos2020response,
  title={Response and sensitivity using Markov chains},
  author={Santos Guti{\'e}rrez, Manuel and Lucarini, Valerio},
  journal={Journal of Statistical Physics},
  volume={179},
  number={5},
  pages={1572--1593},
  year={2020},
  publisher={Springer},
doi={10.1007/s10955-020-02504-4}
}

@article{lucarini2016response,
  title={Response operators for Markov processes in a finite state space: radius of convergence and link to the response theory for Axiom A systems},
  author={Lucarini, Valerio},
  journal={Journal of Statistical Physics},
  volume={162},
  pages={312--333},
  year={2016},
  publisher={Springer},
doi={10.1007/s10955-015-1409-4}
}

@article{shiraishi2021optimal,
  title={Optimal thermodynamic uncertainty relation in Markov jump processes},
  author={Shiraishi, Naoto},
  journal={Journal of Statistical Physics},
  volume={185},
  number={3},
  pages={19},
  year={2021},
  publisher={Springer},
doi={10.1007/s10955-021-02829-8}
}

@article{van2023thermodynamic,
  title={Thermodynamic unification of optimal transport: Thermodynamic uncertainty relation, minimum dissipation, and thermodynamic speed limits},
  author={Van Vu, Tan and Saito, Keiji},
  journal={Physical Review X},
  volume={13},
  number={1},
  pages={011013},
  year={2023},
  publisher={APS},
doi={10.1103/PhysRevX.13.011013}
}

@article{floyd2024limits,
  title={Limits on the computational expressivity of non-equilibrium biophysical processes},
  author={Floyd, Carlos and Dinner, Aaron R and Murugan, Arvind and Vaikuntanathan, Suriyanarayanan},
  journal={Nature Communications},
  volume={16},
  number={1},
  pages={7184},
  year={2025},
  publisher={Nature Publishing Group UK London},
doi={10.1038/s41467-025-61873-0}
}

@article{frezzato2024steady,
  title={Steady-state probabilities for Markov jump processes in terms of powers of the transition rate matrix},
  author={Frezzato, Diego},
  journal={The Journal of Chemical Physics},
  volume={160},
  number={23},
pages={234111},
  year={2024},
  publisher={AIP Publishing},
doi={10.1063/5.0217202}
}

@article{di2018kinetic,
  title={Kinetic uncertainty relation},
  author={Di Terlizzi, Ivan and Baiesi, Marco},
  journal={Journal of Physics A: Mathematical and Theoretical},
  volume={52},
  number={2},
  pages={02LT03},
  year={2018},
  publisher={IOP Publishing},
  doi = {10.1088/1751-8121/aaee34}
}

@article{vu2020entropy,
  title={Entropy production estimation with optimal current},
  author={Van Vu, Tan and Vo, Van Tuan and Hasegawa, Yoshihiko},
  journal={Phys. Rev. E},
  volume={101},
  number={4},
  pages={042138},
  year={2020},
  publisher={APS},
doi = {10.1103/PhysRevE.101.042138}
}

@article{aslyamov2024general,
  title = {General Theory of Static Response for Markov Jump Processes},
  author = {Aslyamov, Timur and Esposito, Massimiliano},
  journal = {Phys. Rev. Lett.},
  volume = {133},
  issue = {10},
  pages = {107103},
  numpages = {7},
  year = {2024},
  month = {Sep},
  publisher = {American Physical Society},
  doi = {10.1103/PhysRevLett.133.107103},
  url = {https://link.aps.org/doi/10.1103/PhysRevLett.133.107103}
}

@article{gao2022thermodynamic,
  title={Thermodynamic constraints on the nonequilibrium response of one-dimensional diffusions},
  author={Gao, Qi and Chun, Hyun-Myung and Horowitz, Jordan M},
  journal={Phys. Rev. E},
  volume={105},
  number={1},
  pages={L012102},
  year={2022},
  publisher={APS},
doi={10.1103/PhysRevE.105.L012102}
}

@article{zheng2024information,
  title = {Universal response inequalities beyond steady states via trajectory information geometry},
  author = {Zheng, Jiming and Lu, Zhiyue},
  journal = {Phys. Rev. E},
  volume = {112},
  issue = {1},
  pages = {L012103},
  numpages = {6},
  year = {2025},
  month = {Jul},
  publisher = {American Physical Society},
  doi = {10.1103/scg2-qkxv},
  url = {https://link.aps.org/doi/10.1103/scg2-qkxv}
}

@article{harunari2024mutual,
  title = {Mutual Linearity of Nonequilibrium Network Currents},
  author = {Harunari, Pedro E. and Dal Cengio, Sara and Lecomte, Vivien and Polettini, Matteo},
  journal = {Phys. Rev. Lett.},
  volume = {133},
  issue = {4},
  pages = {047401},
  numpages = {8},
  year = {2024},
  month = {Jul},
  publisher = {American Physical Society},
  doi = {10.1103/PhysRevLett.133.047401},
  url = {https://link.aps.org/doi/10.1103/PhysRevLett.133.047401}
}

@article{floyd2024learning,
  title={Learning to control non-equilibrium dynamics using local imperfect gradients},
  author={Floyd, Carlos and Dinner, Aaron R and Vaikuntanathan, Suriyanarayanan},
  journal={arXiv preprint},
  year={2024},
doi={10.48550/arXiv.2404.03798}
}

@article{gao2024thermodynamic,
  title={Thermodynamic constraints on kinetic perturbations of homogeneous driven diffusions},
  author={Gao, Qi and Chun, Hyun-Myung and Horowitz, Jordan M},
  journal={Europhys. Lett.},
  year={2024},
volume={146},
pages={31001},
doi={10.1209/0295-5075/ad40cd}
}

@article{gabriela2023topologically,
  title = {Topologically constrained fluctuations and thermodynamics regulate nonequilibrium response},
  author = {Fernandes Martins, Gabriela and Horowitz, Jordan M.},
  journal = {Phys. Rev. E},
  volume = {108},
  issue = {4},
  pages = {044113},
  numpages = {21},
  year = {2023},
  month = {Oct},
  publisher = {American Physical Society},
  doi = {10.1103/PhysRevE.108.044113},
  url = {https://link.aps.org/doi/10.1103/PhysRevE.108.044113}
}

@article{aslyamov2024nonequilibrium,
  title = {Nonequilibrium Response for Markov Jump Processes: Exact Results and Tight Bounds},
  author = {Aslyamov, Timur and Esposito, Massimiliano},
  journal = {Phys. Rev. Lett.},
  volume = {132},
  issue = {3},
  pages = {037101},
  numpages = {6},
  year = {2024},
  month = {Jan},
  publisher = {American Physical Society},
  doi = {10.1103/PhysRevLett.132.037101}
}

@article{aslyamov2024frr,
  title = {Nonequilibrium Fluctuation-Response Relations: From Identities to Bounds},
  author = {Aslyamov, Timur and Ptaszy\ifmmode \acute{n}\else \'{n}\fi{}ski, Krzysztof and Esposito, Massimiliano},
  journal = {Phys. Rev. Lett.},
  volume = {134},
  issue = {15},
  pages = {157101},
  numpages = {9},
  year = {2025},
  month = {Apr},
  publisher = {American Physical Society},
  doi = {10.1103/PhysRevLett.134.157101},
  url = {https://link.aps.org/doi/10.1103/PhysRevLett.134.157101}
}

@article{mallory2020kinetic,
  title={Kinetic control of stationary flux ratios for a wide range of biochemical processes},
  author={Mallory, Joel D and Kolomeisky, Anatoly B and Igoshin, Oleg A},
  journal={Proceedings of the National Academy of Sciences},
  volume={117},
  number={16},
  pages={8884--8889},
  year={2020},
  publisher={National Acad Sciences},
doi={10.1073/pnas.1920873117}
}

@article{govern2014energy,
  title={Energy dissipation and noise correlations in biochemical sensing},
  author={Govern, Christopher C and ten Wolde, Pieter Rein},
  journal={Phys. Rev. Lett.},
  volume={113},
  number={25},
  pages={258102},
  year={2014},
  publisher={APS},
doi={10.1103/PhysRevLett.113.258102}
}

@article{agarwal1972fluctuation,
  title={Fluctuation-dissipation theorems for systems in non-thermal equilibrium and applications},
  author={Agarwal, Girish Saran},
  journal={Zeitschrift f{\"u}r Physik A Hadrons and nuclei},
  volume={252},
  number={1},
  pages={25--38},
  year={1972},
  publisher={Springer},
doi={10.1007/BF01391621}
}

@article{owen2020universal,
  title={Universal thermodynamic bounds on nonequilibrium response with biochemical applications},
  author={Owen, Jeremy A and Gingrich, Todd R and Horowitz, Jordan M},
  journal={Phys. Rev. X},
  volume={10},
  number={1},
  pages={011066},
  year={2020},
  publisher={APS},
doi={10.1103/PhysRevX.10.011066}
}

@article{prost2009generalized,
  title={Generalized fluctuation-dissipation theorem for steady-state systems},
  author={Prost, Jacques and Joanny, J-F and Parrondo, Juan MR},
  journal={Phys. Rev. Lett.},
  volume={103},
  number={9},
  pages={090601},
  year={2009},
  publisher={APS},
doi={10.1103/PhysRevLett.103.090601}
}

@article{seifert2010fluctuation,
  title={Fluctuation-dissipation theorem in nonequilibrium steady states},
  author={Seifert, Udo and Speck, Thomas},
  journal={Europhys. Lett.},
  volume={89},
  number={1},
  pages={10007},
  year={2010},
  publisher={IOP Publishing},
doi={10.1209/0295-5075/89/10007}
}

@article{baiesi2009fluctuations,
  title={Fluctuations and response of nonequilibrium states},
  author={Baiesi, Marco and Maes, Christian and Wynants, Bram},
  journal={Phys. Rev. Lett.},
  volume={103},
  number={1},
  pages={010602},
  year={2009},
  publisher={APS},
doi={10.1103/PhysRevLett.103.010602}
}

@article{altaner2016fluctuation,
  title={Fluctuation-dissipation relations far from equilibrium},
  author={Altaner, Bernhard and Polettini, Matteo and Esposito, Massimiliano},
  journal={Phys. Rev. Lett.},
  volume={117},
  number={18},
  pages={180601},
  year={2016},
  publisher={APS},
doi={10.1103/PhysRevLett.117.180601}
}

@article{kubo1966fluctuation,
  title={The fluctuation-dissipation theorem},
  author={Kubo, Rep},
  journal={Rep. Prog. Phys.},
  volume={29},
  number={1},
  pages={255},
  year={1966},
  publisher={IOP Publishing},
doi={10.1088/0034-4885/29/1/306}
}

@article{forastiere2022linear,
  title={Linear stochastic thermodynamics},
  author={Forastiere, Danilo and Rao, Riccardo and Esposito, Massimiliano},
  journal={New Journal of Physics},
  volume={24},
  number={8},
  pages={083021},
  year={2022},
  publisher={IOP Publishing},
doi={10.1088/1367-2630/ac836b}
}

@article{rao2018conservation,
  title={Conservation laws shape dissipation},
  author={Rao, Riccardo and Esposito, Massimiliano},
  journal={New Journal of Physics},
  volume={20},
  number={2},
  pages={023007},
  year={2018},
  publisher={IOP Publishing},
doi={10.1088/1367-2630/aaa15f}
}

@article{owen2023size,
  title={Size limits the sensitivity of kinetic schemes},
  author={Owen, Jeremy A and Horowitz, Jordan M},
  journal={Nature Communications},
  volume={14},
  number={1},
  pages={1280},
  year={2023},
  publisher={Nature Publishing Group UK London},
doi={10.1038/s41467-023-36705-8}
}

@article{marconi2008fluctuation,
  title={Fluctuation--dissipation: response theory in statistical physics},
  author={Marconi, Umberto Marini Bettolo and Puglisi, Andrea and Rondoni, Lamberto and Vulpiani, Angelo},
  journal={Physics reports},
  volume={461},
  number={4-6},
  pages={111--195},
  year={2008},
  publisher={Elsevier},
doi={https://doi.org/10.1016/j.physrep.2008.02.002}
}

@article{barato2015thermodynamic,
  title={Thermodynamic uncertainty relation for biomolecular processes},
  author={Barato, Andre C and Seifert, Udo},
  journal={Phys. Rev. Lett.},
  volume={114},
  number={15},
  pages={158101},
  year={2015},
  publisher={APS},
doi={10.1103/PhysRevLett.114.158101}
}

@article{gingrich2016dissipation,
  title={Dissipation bounds all steady-state current fluctuations},
  author={Gingrich, Todd R and Horowitz, Jordan M and Perunov, Nikolay and England, Jeremy L},
  journal={Phys. Rev. Lett.},
  volume={116},
  number={12},
  pages={120601},
  year={2016},
  publisher={APS},
doi={10.1103/PhysRevLett.116.120601}
}

@article{falasco2019negative,
  title={Negative differential response in chemical reactions},
  author={Falasco, Gianmaria and Cossetto, Tommaso and Penocchio, Emanuele and Esposito, Massimiliano},
  journal={New Journal of Physics},
  volume={21},
  number={7},
  pages={073005},
  year={2019},
  publisher={IOP Publishing},
doi={10.1088/1367-2630/ab28be}
}

@article{falasco2023macroscopic,
  title = {Macroscopic stochastic thermodynamics},
  author = {Falasco, Gianmaria and Esposito, Massimiliano},
  journal = {Rev. Mod. Phys.},
  volume = {97},
  issue = {1},
  pages = {015002},
  numpages = {46},
  year = {2025},
  month = {Jan},
  publisher = {American Physical Society},
  doi = {10.1103/RevModPhys.97.015002},
  url = {https://link.aps.org/doi/10.1103/RevModPhys.97.015002}
}

@article{chun2023trade,
  title={Trade-offs between number fluctuations and response in nonequilibrium chemical reaction networks},
  author={Chun, Hyun-Myung and Horowitz, Jordan M},
  journal={The Journal of Chemical Physics},
  volume={158},
  number={17},
pages={174115},
  year={2023},
  publisher={AIP Publishing},
doi={https://doi.org/10.1063/5.0148662}
}

@article{dechant2020fluctuation,
  title={Fluctuation--response inequality out of equilibrium},
  author={Dechant, Andreas and Sasa, Shin-ichi},
  journal={Proceedings of the National Academy of Sciences},
  volume={117},
  number={12},
  pages={6430--6436},
  year={2020},
  publisher={National Acad Sciences},
doi={10.1073/pnas.1918386117}
}

@article{barato2015dispersion,
  title = {Dispersion for two classes of random variables: General theory and application to inference of an external ligand concentration by a cell},
  author = {Barato, Andre C. and Seifert, Udo},
  journal = {Phys. Rev. E},
  volume = {92},
  issue = {3},
  pages = {032127},
  numpages = {9},
  year = {2015},
  month = {Sep},
  publisher = {American Physical Society},
  doi = {10.1103/PhysRevE.92.032127},
  url = {https://link.aps.org/doi/10.1103/PhysRevE.92.032127}
}

@article{landi2023current,
  title = "{Current Fluctuations in Open Quantum Systems: Bridging the Gap Between Quantum Continuous Measurements and Full Counting Statistics}",
  author = {Landi, Gabriel T. and Kewming, Michael J. and Mitchison, Mark T. and Potts, Patrick P.},
  journal = {PRX Quantum},
  volume = {5},
  issue = {2},
  pages = {020201},
  numpages = {86},
  year = {2024},
  month = {Apr},
  publisher = {American Physical Society},
  doi = {10.1103/PRXQuantum.5.020201},
  url = {https://link.aps.org/doi/10.1103/PRXQuantum.5.020201}
}

@article{gustavsson2006counting,
  title = "{Counting Statistics of Single Electron Transport in a Quantum Dot}",
  author = {Gustavsson, S. and Leturcq, R. and Simovi\ifmmode \check{c}\else \v{c}\fi{}, B. and Schleser, R. and Ihn, T. and Studerus, P. and Ensslin, K. and Driscoll, D. C. and Gossard, A. C.},
  journal = {Phys. Rev. Lett.},
  volume = {96},
  issue = {7},
  pages = {076605},
  numpages = {4},
  year = {2006},
  month = {Feb},
  publisher = {American Physical Society},
  doi = {10.1103/PhysRevLett.96.076605},
  url = {https://link.aps.org/doi/10.1103/PhysRevLett.96.076605}
}

@article{gustavsson2009electron,
  title={Electron counting in quantum dots},
  author={Gustavsson, Simon and Leturcq, R and Studer, M and Shorubalko, I and Ihn, T and Ensslin, K and Driscoll, DC and Gossard, AC},
  journal={Surface Science Reports},
  volume={64},
  number={6},
  pages={191--232},
  year={2009},
  publisher={Elsevier},
doi={10.1016/j.surfrep.2009.02.001}
}

@article{pietzonka2016universal,
  title = {Universal bounds on current fluctuations},
  author = {Pietzonka, Patrick and Barato, Andre C. and Seifert, Udo},
  journal = {Phys. Rev. E},
  volume = {93},
  issue = {5},
  pages = {052145},
  numpages = {16},
  year = {2016},
  month = {May},
  publisher = {American Physical Society},
  doi = {10.1103/PhysRevE.93.052145},
  url = {https://link.aps.org/doi/10.1103/PhysRevE.93.052145}
}

@article{pietzonka2017finite,
  title = {Finite-time generalization of the thermodynamic uncertainty relation},
  author = {Pietzonka, Patrick and Ritort, Felix and Seifert, Udo},
  journal = {Phys. Rev. E},
  volume = {96},
  issue = {1},
  pages = {012101},
  numpages = {6},
  year = {2017},
  month = {Jul},
  publisher = {American Physical Society},
  doi = {10.1103/PhysRevE.96.012101},
  url = {https://link.aps.org/doi/10.1103/PhysRevE.96.012101}
}

@article{horowitz2017proof,
  title = {Proof of the finite-time thermodynamic uncertainty relation for steady-state currents},
  author = {Horowitz, Jordan M. and Gingrich, Todd R.},
  journal = {Phys. Rev. E},
  volume = {96},
  issue = {2},
  pages = {020103},
  numpages = {3},
  year = {2017},
  month = {Aug},
  publisher = {American Physical Society},
  doi = {10.1103/PhysRevE.96.020103},
  url = {https://link.aps.org/doi/10.1103/PhysRevE.96.020103}
}

@article{falasco2020unifying,
  title={Unifying thermodynamic uncertainty relations},
  author={Falasco, Gianmaria and Esposito, Massimiliano and Delvenne, Jean-Charles},
  journal={New Journal of Physics},
  volume={22},
  number={5},
  pages={053046},
  year={2020},
  publisher={IOP Publishing},
doi={10.1088/1367-2630/ab8679}
}

@article{horowitz2020thermodynamic,
  title={Thermodynamic uncertainty relations constrain non-equilibrium fluctuations},
  author={Horowitz, Jordan M and Gingrich, Todd R},
  journal={Nature Physics},
  volume={16},
  number={1},
  pages={15--20},
  year={2020},
  publisher={Nature Publishing Group UK London},
doi={10.1038/s41567-019-0702-6}
}

@article{seifert2012stochastic,
  title={Stochastic thermodynamics, fluctuation theorems and molecular machines},
  author={Seifert, Udo},
  journal={Rep. Prog. Phys.},
  volume={75},
  number={12},
  pages={126001},
  year={2012},
  publisher={IOP Publishing},
doi={10.1088/0034-4885/75/12/126001}
}

@article{falasco2022beyond,
  title={Beyond thermodynamic uncertainty relations: Nonlinear response, error-dissipation trade-offs, and speed limits},
  author={Falasco, Gianmaria and Esposito, Massimiliano and Delvenne, Jean-Charles},
  journal={Journal of Physics A: Mathematical and Theoretical},
  volume={55},
  number={12},
  pages={124002},
  year={2022},
  publisher={IOP Publishing},
doi={10.1088/1751-8121/ac52e2}
}

@misc{crook2018drazin,
  author      = "Crooks, Gavin E.",
  title      = "On the {D}razin inverse of the rate matrix",
note="Technical note",
  year        = "2018",
url="https://threeplusone.com/pubs/drazin03/"
}

@misc{macieszczak2024occupation,
      title={Occupation Uncertainty Relations}, 
      author={Katarzyna Macieszczak},
      year={2024},
      eprint={2411.15118},
      archivePrefix={arXiv},
      primaryClass={cond-mat.stat-mech}
}

@article{berg1977physics,
  title={Physics of chemoreception},
  author={Berg, Howard C and Purcell, Edward M},
  journal={Biophysical J.},
  volume={20},
  number={2},
  pages={193--219},
  year={1977},
  publisher={Elsevier},
doi={10.1016/S0006-3495(77)85544-6}
}

@article{koyuk2020thermodynamic,
  title = {Thermodynamic Uncertainty Relation for Time-Dependent Driving},
  author = {Koyuk, Timur and Seifert, Udo},
  journal = {Phys. Rev. Lett.},
  volume = {125},
  issue = {26},
  pages = {260604},
  numpages = {6},
  year = {2020},
  month = {Dec},
  publisher = {American Physical Society},
  doi = {10.1103/PhysRevLett.125.260604},
  url = {https://link.aps.org/doi/10.1103/PhysRevLett.125.260604}
}

@article{lapolla2020spectral,
  title = {Spectral theory of fluctuations in time-average statistical mechanics of reversible and driven systems},
  author = {Lapolla, Alessio and Hartich, David and Godec, A},
  journal = {Phys. Rev. Res.},
  volume = {2},
  issue = {4},
  pages = {043084},
  numpages = {17},
  year = {2020},
  month = {Oct},
  publisher = {American Physical Society},
  doi = {10.1103/PhysRevResearch.2.043084},
  url = {https://link.aps.org/doi/10.1103/PhysRevResearch.2.043084}
}

@article{gopich2012theory,
  title={Theory of the energy transfer efficiency and fluorescence lifetime distribution in single-molecule FRET},
  author={Gopich, Irina V and Szabo, Attila},
  journal={Proceedings of the National Academy of Sciences},
  volume={109},
  number={20},
  pages={7747--7752},
  year={2012},
  publisher={National Acad Sciences},
doi={https://doi.org/10.1073/pnas.1205120109}
}

@article{dhar1999residence,
  title = {Residence time distribution for a class of Gaussian Markov processes},
  author = {Dhar, Abhishek and Majumdar, Satya N.},
  journal = {Phys. Rev. E},
  volume = {59},
  issue = {6},
  pages = {6413--6418},
  numpages = {0},
  year = {1999},
  month = {Jun},
  publisher = {American Physical Society},
  doi = {10.1103/PhysRevE.59.6413},
  url = {https://link.aps.org/doi/10.1103/PhysRevE.59.6413}
}

@incollection{majumdar2007brownian,
  title={Brownian functionals in physics and computer science},
  author={Majumdar, Satya N},
  booktitle={The legacy of Albert Einstein: A collection of essays in celebration of the year of physics},
  pages={93--129},
  year={2007},
  publisher={World Scientific}
}

@article{lapolla2018unfolding,
  title={Unfolding tagged particle histories in single-file diffusion: exact single-and two-tag local times beyond large deviation theory},
  author={Lapolla, Alessio and Godec, Alja{\v{z}}},
  journal={New Journal of Physics},
  volume={20},
  number={11},
  pages={113021},
  year={2018},
  publisher={IOP Publishing},
doi={10.1088/1367-2630/aaea1b}
}

@article{lapolla2019manifestations,
  title={Manifestations of projection-induced memory: General theory and the tilted single file},
  author={Lapolla, Alessio and Godec, A},
  journal={Frontiers in Physics},
  volume={7},
  pages={182},
  year={2019},
  publisher={Frontiers Media SA},
doi={10.3389/fphy.2019.00182}
}

@article{landauer1962fluctuations,
  title={Fluctuations in bistable tunnel diode circuits},
  author={Landauer, Rolf},
  journal={J. App. Phys.},
  volume={33},
  number={7},
  pages={2209},
  year={1962},
  publisher={American Institute of Physics},
doi={10.1063/1.1728929}
}

@article{hanggi1982stochastic,
  title={Stochastic processes: {T}ime evolution, symmetries and linear response},
  author={H{\"a}nggi, Peter and Thomas, Harry},
  journal={Phys. Rep.},
  volume={88},
  number={4},
  pages={207},
  year={1982},
  publisher={Elsevier},
doi={10.1016/0370-1573(82)90045-X}
}

@article{herpich2020njp,
	doi = {10.1088/1367-2630/ab882f},
	url = {https://doi.org/10.1088%2F1367-2630%2Fab882f},
	year = {2020},
	month = {jun},
	publisher = {{IOP} Publishing},
	volume = {22},
	number = {6},
	pages = {063005},
	author = {Tim Herpich and Tommaso Cossetto and Gianmaria Falasco and Massimiliano Esposito},
	title = {Stochastic thermodynamics of all-to-all interacting many-body systems},
	journal = {New J. Phys.}
}

@article{meibohm2022finite,
  title={{Finite-Time Dynamical Phase Transition in Nonequilibrium Relaxation}},
  author={Meibohm, Jan and Esposito, Massimiliano},
  journal={Phys. Rev. Lett.},
  volume={128},
  number={11},
  pages={110603},
  year={2022},
  publisher={APS},
doi={10.1103/PhysRevLett.128.110603}
}

@article{meibohm2023landau,
  title={Landau theory for finite-time dynamical phase transitions},
  author={Meibohm, Jan and Esposito, Massimiliano},
  journal={New J. Phys.},
  volume={25},
  number={2},
  pages={023034},
  year={2023},
  publisher={IOP Publishing},
doi={10.1088/1367-2630/acbc41}
}

@article{ptaszynski2024critical,
  title = {{Critical heat current fluctuations in Curie-Weiss model in and out of equilibrium}},
  author = {Ptaszy\ifmmode \acute{n}\else \'{n}\fi{}ski, Krzysztof and Esposito, Massimiliano},
  journal = {Phys. Rev. E},
  volume = {111},
  issue = {3},
  pages = {034125},
  numpages = {15},
  year = {2025},
  month = {Mar},
  publisher = {American Physical Society},
  doi = {10.1103/PhysRevE.111.034125},
  url = {https://link.aps.org/doi/10.1103/PhysRevE.111.034125}
}

@article{vroylandt2018collective,
  title={Collective effects enhancing power and efficiency},
  author={Vroylandt, Hadrien and Esposito, Massimiliano and Verley, Gatien},
  journal={Europhys. Lett.},
  volume={120},
  number={3},
  pages={30009},
  year={2017},
  publisher={IOP Publishing},
doi={10.1209/0295-5075/120/30009}
}

@article{vroylandt2020efficiency,
  title = "{Efficiency Fluctuations of Stochastic Machines Undergoing a Phase Transition}",
  author = {Vroylandt, Hadrien and Esposito, Massimiliano and Verley, Gatien},
  journal = {Phys. Rev. Lett.},
  volume = {124},
  issue = {25},
  pages = {250603},
  numpages = {6},
  year = {2020},
  month = {Jun},
  publisher = {American Physical Society},
  doi = {10.1103/PhysRevLett.124.250603},
  url = {https://link.aps.org/doi/10.1103/PhysRevLett.124.250603}
}

@article{ge2009thermodynamic,
	title={{Thermodynamic Limit of a Nonequilibrium Steady State: Maxwell-Type Construction for a Bistable Biochemical System}},
	author={Ge, Hao and Qian, Hong},
	journal={Phys. Rev. Lett.},
	volume={103},
	number={14},
	pages={148103},
	year={2009},
	publisher={APS},
	doi={10.1103/PhysRevLett.103.148103}
}

@article{vellela2009stochastic,
  title={Stochastic dynamics and non-equilibrium thermodynamics of a bistable chemical system: the Schl{\"o}gl model revisited},
  author={Vellela, Melissa and Qian, Hong},
  journal={Journal of The Royal Society Interface},
  volume={6},
  number={39},
  pages={925--940},
  year={2009},
  publisher={The Royal Society},
doi={10.1098/rsif.2008.0476}
}

@article{schlogl1972chemical,
  title={Chemical reaction models for non-equilibrium phase transitions},
  author={Schl{\"o}gl, Friedrich},
  journal={Z. Physik},
  volume={253},
  number={2},
  pages={147--161},
  year={1972},
  publisher={Springer},
doi={10.1007/BF01379769}
}

@article{nieddu2022characterizing,
  title={Characterizing outbreak vulnerability in a stochastic SIS model with an external disease reservoir},
  author={Nieddu, Garrett T and Forgoston, Eric and Billings, Lora},
  journal={Journal of the Royal Society Interface},
  volume={19},
  number={192},
  pages={20220253},
  year={2022},
  publisher={The Royal Society},
doi={10.1098/rsif.2022.0253}
}

@article{ray2017dispersion,
  title={Dispersion of the time spent in a state: general expression for unicyclic model and dissipation-less precision},
  author={Ray, Somrita and Barato, Andre C},
  journal={Journal of Physics A: Mathematical and Theoretical},
  volume={50},
  number={35},
  pages={355001},
  year={2017},
  publisher={IOP Publishing},
doi={10.1088/1751-8121/aa7f7a}
}

@article{bakewell2023general,
  title = {General Upper Bounds on Fluctuations of Trajectory Observables},
  author = {Bakewell-Smith, George and Girotti, Federico and Gu\ifmmode \mbox{\c{t}}\else \c{t}\fi{}\ifmmode \u{a}\else \u{a}\fi{}, M and Garrahan, Juan P.},
  journal = {Phys. Rev. Lett.},
  volume = {131},
  issue = {19},
  pages = {197101},
  numpages = {6},
  year = {2023},
  month = {Nov},
  publisher = {American Physical Society},
  doi = {10.1103/PhysRevLett.131.197101},
  url = {https://link.aps.org/doi/10.1103/PhysRevLett.131.197101}
}

@article{jarzynski1997nonequilibrium,
  title = {Nonequilibrium Equality for Free Energy Differences},
  author = {Jarzynski, C.},
  journal = {Phys. Rev. Lett.},
  volume = {78},
  issue = {14},
  pages = {2690--2693},
  numpages = {0},
  year = {1997},
  month = {Apr},
  publisher = {American Physical Society},
  doi = {10.1103/PhysRevLett.78.2690},
  url = {https://link.aps.org/doi/10.1103/PhysRevLett.78.2690}
}

@article{crooks1999entropy,
  title = {Entropy production fluctuation theorem and the nonequilibrium work relation for free energy differences},
  author = {Crooks, Gavin E.},
  journal = {Phys. Rev. E},
  volume = {60},
  issue = {3},
  pages = {2721--2726},
  numpages = {0},
  year = {1999},
  month = {Sep},
  publisher = {American Physical Society},
  doi = {10.1103/PhysRevE.60.2721},
  url = {https://link.aps.org/doi/10.1103/PhysRevE.60.2721}
}

@article{esposito2009nonequilibrium,
  title = {Nonequilibrium fluctuations, fluctuation theorems, and counting statistics in quantum systems},
  author = {Esposito, Massimiliano and Harbola, Upendra and Mukamel, Shaul},
  journal = {Rev. Mod. Phys.},
  volume = {81},
  issue = {4},
  pages = {1665--1702},
  numpages = {0},
  year = {2009},
  month = {Dec},
  publisher = {American Physical Society},
  doi = {10.1103/RevModPhys.81.1665},
  url = {https://link.aps.org/doi/10.1103/RevModPhys.81.1665}
}

@article{bialek2005physical,
  title={Physical limits to biochemical signaling},
  author={Bialek, William and Setayeshgar, Sima},
  journal={Proceedings of the National Academy of Sciences},
  volume={102},
  number={29},
  pages={10040--10045},
  year={2005},
  publisher={National Acad Sciences},
doi={10.1073/pnas.0504321102}
}

@article{endres2008accuracy,
  title={Accuracy of direct gradient sensing by single cells},
  author={Endres, Robert G and Wingreen, Ned S},
  journal={Proceedings of the National Academy of Sciences},
  volume={105},
  number={41},
  pages={15749--15754},
  year={2008},
  publisher={National Acad Sciences},
doi={10.1073/pnas.0804688105}
}

@article{govern2012fundamental,
  title = {Fundamental Limits on Sensing Chemical Concentrations with Linear Biochemical Networks},
  author = {Govern, Christopher C. and ten Wolde, Pieter Rein},
  journal = {Phys. Rev. Lett.},
  volume = {109},
  issue = {21},
  pages = {218103},
  numpages = {5},
  year = {2012},
  month = {Nov},
  publisher = {American Physical Society},
  doi = {10.1103/PhysRevLett.109.218103},
  url = {https://link.aps.org/doi/10.1103/PhysRevLett.109.218103}
}

@article{kaizu2014berg,
  title={The berg-purcell limit revisited},
  author={Kaizu, Kazunari and De Ronde, Wiet and Paijmans, Joris and Takahashi, Koichi and Tostevin, Filipe and Ten Wolde, Pieter Rein},
  journal={Biophysical J.},
  volume={106},
  number={4},
  pages={976--985},
  year={2014},
  publisher={Elsevier},
doi={10.1016/j.bpj.2013.12.030}
}

@article{lang2014thermodynamics,
  title = {Thermodynamics of Statistical Inference by Cells},
  author = {Lang, Alex H. and Fisher, Charles K. and Mora, Thierry and Mehta, Pankaj},
  journal = {Phys. Rev. Lett.},
  volume = {113},
  issue = {14},
  pages = {148103},
  numpages = {5},
  year = {2014},
  month = {Oct},
  publisher = {American Physical Society},
  doi = {10.1103/PhysRevLett.113.148103},
  url = {https://link.aps.org/doi/10.1103/PhysRevLett.113.148103}
}

@article{palo2006calculation,
  title={Calculation of photon-count number distributions via master equations},
  author={Palo, Kaupo and Mets, {\"U}lo and Loorits, Vello and Kask, Peet},
  journal={Biophysical J.},
  volume={90},
  number={6},
  pages={2179--2191},
  year={2006},
  publisher={Elsevier},
doi={10.1529/biophysj.105.066084}}

@article{tesser2024out,
  title = {Out-of-Equilibrium Fluctuation-Dissipation Bounds},
  author = {Tesser, Ludovico and Splettstoesser, Janine},
  journal = {Phys. Rev. Lett.},
  volume = {132},
  issue = {18},
  pages = {186304},
  numpages = {7},
  year = {2024},
  month = {May},
  publisher = {American Physical Society},
  doi = {10.1103/PhysRevLett.132.186304},
  url = {https://link.aps.org/doi/10.1103/PhysRevLett.132.186304}
}

@article{levy1940certains,
  title={Sur certains processus stochastiques homog{\`e}nes},
  author={L{\'e}vy, Paul},
  journal={Compositio Mathematica},
  volume={7},
  pages={283--339},
  year={1940}
}

@article{majumdar2002local,
  title = {Local and Occupation Time of a Particle Diffusing in a Random Medium},
  author = {Majumdar, Satya  N. and Comtet, Alain},
  journal = {Phys. Rev. Lett.},
  volume = {89},
  issue = {6},
  pages = {060601},
  numpages = {4},
  year = {2002},
  month = {Jul},
  publisher = {American Physical Society},
  doi = {10.1103/PhysRevLett.89.060601},
  url = {https://link.aps.org/doi/10.1103/PhysRevLett.89.060601}
}

@article{godreche2001statistics,
  title={Statistics of the occupation time of renewal processes},
  author={Godreche, C and Luck, JM1853425},
  journal={Journal of Statistical Physics},
  volume={104},
  pages={489--524},
  year={2001},
  publisher={Springer},
doi={10.1023/A:1010364003250}
}

@article{margolin2005nonergodicity,
  title = {Nonergodicity of Blinking Nanocrystals and Other L\'evy-Walk Processes},
  author = {Margolin, G. and Barkai, E.},
  journal = {Phys. Rev. Lett.},
  volume = {94},
  issue = {8},
  pages = {080601},
  numpages = {4},
  year = {2005},
  month = {Mar},
  publisher = {American Physical Society},
  doi = {10.1103/PhysRevLett.94.080601},
  url = {https://link.aps.org/doi/10.1103/PhysRevLett.94.080601}
}

@article{ramesh2024arcine,
  title = {Arcsine Laws of Light},
  author = {Ramesh, V. G. and Peters, K. J. H. and Rodriguez, S. R. K.},
  journal = {Phys. Rev. Lett.},
  volume = {132},
  issue = {13},
  pages = {133801},
  numpages = {7},
  year = {2024},
  month = {Mar},
  publisher = {American Physical Society},
  doi = {10.1103/PhysRevLett.132.133801},
  url = {https://link.aps.org/doi/10.1103/PhysRevLett.132.133801}
}

@article{singh2019generalised,
  title={Generalised ‘Arcsine’laws for run-and-tumble particle in one dimension},
  author={Singh, Prashant and Kundu, Anupam},
  journal={Journal of Statistical Mechanics: Theory and Experiment},
  volume={2019},
  number={8},
  pages={083205},
  year={2019},
  publisher={IOP Publishing},
doi={10.1088/1742-5468/ab3283}
}

@article{bresloff2020occupation,
  title = {Occupation time of a run-and-tumble particle with resetting},
  author = {Bressloff, Paul C.},
  journal = {Phys. Rev. E},
  volume = {102},
  issue = {4},
  pages = {042135},
  numpages = {12},
  year = {2020},
  month = {Oct},
  publisher = {American Physical Society},
  doi = {10.1103/PhysRevE.102.042135},
  url = {https://link.aps.org/doi/10.1103/PhysRevE.102.042135}
}

@article{khodabandehlou2024affine,
  title={Affine relationships between steady currents},
  author={Khodabandehlou, Faezeh and Maes, Christian and Neto{\v{c}}n{\`y}, Karel},
  journal={Journal of Physics A: Mathematical and Theoretical},
  volume={58},
  number={15},
  pages={155002},
  year={2025},
  publisher={IOP Publishing},
doi={10.1088/1751-8121/adc8ea}
}

@article{utsumi2007full,
  title = {Full counting statistics for the number of electrons in a quantum dot},
  author = {Utsumi, Yasuhiro},
  journal = {Phys. Rev. B},
  volume = {75},
  issue = {3},
  pages = {035333},
  numpages = {10},
  year = {2007},
  month = {Jan},
  publisher = {American Physical Society},
  doi = {10.1103/PhysRevB.75.035333},
  url = {https://link.aps.org/doi/10.1103/PhysRevB.75.035333}
}

@article{utsumi2007fullb,
  title={Full counting statistics for charge inside a quantum dot},
  author={Utsumi, Yasuhiro},
  journal={Physica E: Low-dimensional Systems and Nanostructures},
  volume={40},
  number={2},
  pages={355--358},
  year={2007},
  publisher={Elsevier},
doi={10.1016/j.physe.2007.06.053}
}

@article{utsumi2008full,
  title={Full counting statistics for electron number in quantum dots},
  author={Utsumi, Yasuhiro and Golubev, Dmitri S and Sch{\"o}n, Gerd},
  journal={physica status solidi c},
  volume={5},
  number={1},
  pages={154--157},
  year={2008},
  publisher={Wiley Online Library},
doi={10.1002/pssc.200776596}
}

@article{zheng2025spatial,
  title={Spatial Correlation Unifies Nonequilibrium Response Theory for Arbitrary Markov Jump Processes},
  author={Zheng, Jiming and Lu, Zhiyue},
  journal={arXiv preprint arXiv:2501.01050},
  year={2025},
doi = {10.48550/arXiv.2501.01050}
}

@article{bao2024nonequilibrium,
  title={Nonequilibrium Response Theory: From Precision Limits to Strong Perturbation},
  author={Bao, Ruicheng and Liang, Shiling},
  journal={arXiv preprint arXiv:2412.19602},
  year={2024},
doi = {10.48550/arXiv.2412.19602}
}

@article{liu2025response,
      title={Response Kinetic Uncertainty Relation for Markovian Open Quantum System}, 
      author={Kangqiao Liu and Jie Gu},
    journal={arXiv preprint arXiv:2501.04895},
      year={2025},
      doi={10.48550/arXiv.2501.04895} 
}

@misc{cengio2025mutual,
      title={Mutual Multilinearity of Nonequilibrium Network Currents}, 
      author={Sara Dal Cengio and Pedro E. Harunari and Vivien Lecomte and Matteo Polettini},
      year={2025},
      eprint={2502.04298},
      archivePrefix={arXiv},
      primaryClass={cond-mat.stat-mech}
}

@misc{ptaszynski2025mixed,
      title={Nonequilibrium fluctuation-response relations for state-current correlations}, 
      author={Krzysztof Ptaszynski and Timur Aslyamov and Massimiliano Esposito},
      year={2025},
      eprint={2506.08877},
      archivePrefix={arXiv},
      primaryClass={cond-mat.stat-mech},
      url={https://arxiv.org/abs/2506.08877}, 
}

@article{monnai2024kinetic,
  title = {Kinetic equality for susceptibility and dynamical activity},
  author = {Monnai, Takaaki},
  journal = {Phys. Rev. E},
  volume = {110},
  issue = {6},
  pages = {L062101},
  numpages = {5},
  year = {2024},
  month = {Dec},
  publisher = {American Physical Society},
  doi = {10.1103/PhysRevE.110.L062101},
  url = {https://link.aps.org/doi/10.1103/PhysRevE.110.L062101}
}

@book{carmichael1999statistical1,
  title={{Statistical Methods in Quantum Optics 1}},
  author={H J Carmichael},
  year={1999},
  publisher={Springer},
address={Berlin},
doi={10.1007/978-3-662-03875-8}
}

@article{novozhilov2006biological,
  title={Biological applications of the theory of birth-and-death processes},
  author={Novozhilov, Artem S and Karev, Georgy P and Koonin, Eugene V},
  journal={Brief. Bioinform.},
  volume={7},
  number={1},
  pages={70--85},
  year={2006},
  publisher={Oxford University Press},
doi={10.1093/bib/bbk006}
}

@article{stegmann2017violation,
  title={Violation of detailed balance for charge-transfer statistics in Coulomb-blockade systems},
  author={Stegmann, Philipp and K{\"o}nig, J{\"u}rgen},
  journal={Phys. Status Solidi B},
  volume={254},
  number={3},
  pages={1600507},
  year={2017},
  publisher={Wiley Online Library},
doi={10.1002/pssb.201600507}
}

@article{drazin1958pseudo,
  title={Pseudo-inverses in associative rings and semigroups},
  author={Drazin, MP},
  journal={Am. Math. Mon.},
  volume={65},
  number={7},
  pages={506--514},
  year={1958},
  publisher={Taylor \& Francis},
doi={10.1080/00029890.1958.11991949}
}
\end{document}